\documentclass[a4paper,fleqn,usenatbib]{mnras}

\usepackage{natbib}
\usepackage{soul}

\usepackage[T1]{fontenc}
\usepackage{ae,aecompl}

\usepackage{todonotes}
\usepackage{outlines}
\usepackage{color}

\usepackage{graphicx}	
\usepackage{amsmath}	
\usepackage{amssymb}	
\usepackage{bm}	        

\def\rpic{r_{\pi, c}}
\newcommand{\angstrom}{\mbox{\normalfont\AA}}

\usepackage{natbib}

\newcommand{\avg}[1]{\left\langle #1 \right\rangle}
\newcommand{\xirs}{\xi_{\mathrm{\tt {LRG-MgII}}}^{s}}
\newcommand{\xirsgg}{\xi_{\mathrm{\tt {LRG-LRG}}}^{s}}

\newcommand{\dd}{\mathrm{d}}
\newcommand{\hmsol}{h^{-1}M_{\odot}}

\newcommand{\mcl}{M_{\mathrm{cl}}}
\newcommand{\mpc}{\mathrm{Mpc}}
\newcommand{\hmpc}{h^{-1}\mathrm{Mpc}}

\newcommand{\hkpc}{h^{-1}\mathrm{kpc}}

\newcommand{\kpc}{\mathrm{kpc}}
\newcommand{\msol}{M_{\odot}}

\newcommand{\kms}{\mathrm{km}\,s^{-1}}
\newcommand{\vlos}{{v}_{\mathrm{los}} }

\newcommand{\siglos}{{\sigma}_{\mathrm{los}} }

\newcommand{\rom}[1]{\uppercase\expandafter{\romannumeral #1\relax}}
\usepackage{newtxtext,newtxmath}

\title[Redshift Space Distortion of MgII Absorbers]{Kinematics of MgII
Absorbers from the Redshift-space Distortion Around Massive Quiescent Galaxies}
\author[Zu 2021]{
Ying  Zu$^{1, 2}$\thanks{E-mail: yingzu@sjtu.edu.cn}
\\
$^{1}$Department of Astronomy, School of Physics and Astronomy, Shanghai Jiao Tong
University, Shanghai 200240, China\\
$^{2}$Shanghai Key Laboratory for Particle Physics and Cosmology, Shanghai Jiao Tong University, Shanghai 200240, China
}

\date{Accepted XXX. Received YYY; in original form ZZZ}

\pubyear{2021}

\begin{document}

\label{firstpage}
\pagerange{\pageref{firstpage}--\pageref{lastpage}}
\maketitle

\begin{abstract}
The kinematics of MgII absorbers is the key to understanding the origin of cool,
metal-enriched gas clouds in the circumgalactic medium of massive quiescent
galaxies. Exploiting the fact that the cloud line-of-sight velocity distribution
is the only unknown for predicting the redshift--space distortion~(RSD) of MgII
absorbers from their 3D real--space distribution around galaxies, we develop a
novel method to infer the cool cloud kinematics from the redshift--space
galaxy--cloud cross--correlation $\xi^{s}$. We measure $\xi^{s}$ for
${\sim}10^4$ MgII absorbers around ${\sim}8{\times}10^5$ CMASS galaxies at
$0.4{<}z{<}0.8$. We discover that $\xi^{s}$ does not exhibit a strong
Fingers-of-God effect, but is heavily truncated at velocity ${\sim}300\,\kms$.
We reconstruct both the redshift and real--space cloud number density
distributions inside haloes, $\xi^{s}_{1h}$ and $\xi_{1h}$, respectively. Thus,
for any model of cloud kinematics, we can predict $\xi^{s}_{1h}$ from the
reconstructed $\xi_{1h}$, and self--consistently compare to the observed
$\xi^{s}_{1h}$. We consider four types of cloud kinematics, including an
isothermal model with a single velocity dispersion, a satellite infall model in
which cool clouds reside in the subhaloes, a cloud accretion model in which
clouds follow the cosmic gas accretion, and a tired wind model in which clouds
originate from the galactic wind--driven bubbles. All the four models provide
statistically good fits to the RSD data, but only the tired wind model can
reproduce the observed truncation by propagating ancient wind bubbles at
${\sim}250\,\kms$ on scales ${\sim}400\,\hkpc$. Our method provides an exciting
path to decoding the dynamical origin of metal absorbers from the RSD
measurements with upcoming spectroscopic surveys.
\end{abstract}




\vspace{1in}
\section{Introduction}
\label{sec:intro}

The circumgalactic space in dark matter haloes is permeated by multi-phase gas
with temperatures ranging from $10^2$ to $10^8$ K and densities between
$10^{-6}$ to $1$ cm$^{-3}$ ~\citep{Tumlinson2017}. One of the most intriguing
puzzles of the circumgalactic medium~(CGM) is the apparent abundance of
cool~(${\sim}10^4$ K), metal-enriched gas clouds~\citep[e.g., detected via the
MgII absorption doublets in rest-frame UV along quasar
sightlines;][]{Bergeron1991, Steidel1992, Steidel1994, Churchill1999}, embedded
in the hot~($T{>}10^6\,$K) gaseous corona~\citep{Spitzer1956, Anderson2013}
surrounding massive quiescent galaxies~\citep{Thom2012, Prochaska2013, Werk2014,
Keeney2017}.  Particularly, it is unclear whether the cool clouds are governed
by random, inflowing, or outflowing motion~\citep{Lan2019}.  In this paper, we
explore the kinematics of MgII absorbers around a large sample of luminous red
galaxies~(LRGs) observed by the Baryon Oscillation Spectroscopic
Survey~\citep[BOSS;][]{Dawson2013} at $z{\sim}0.55$, in hopes of paving the path
for a self--consistent redshift--space distortion~(RSD) modelling of metal
absorbers with upcoming spectroscopic surveys like the DESI~\citep{Levi2019} and
PFS~\citep{Takada2014}.

While the hot corona can be naturally explained by the shock-heating of
inflowing gas to the virial temperature of massive dark matter
haloes~\citep{Birnboim2003, Keres2005, Dekel2009}, the physical origin of the
cool, metal-enriched CGM gas surrounding massive quiescent galaxies remains
largely elusive ~\citep[and references therein]{Chen2017}. This is mainly due to
the lack of galactic winds driven by recent star formation in those systems, and
winds are the dominant means of transporting enriched material into the
CGM~\citep{Oppenheimer2006, Peeples2011, Dave2011, Hummels2013, Ford2014,
Ma2016, Fielding2017, Muratov2017, Fielding2020}. Around star-forming galaxies,
cool gas clouds are commonly seen as blue-shifted absorption and emission in
galactic winds~\citep[see ][and references therein]{Heckman2017, Veilleux2020}.
They could be dragged into the CGM via entrainment in a fast-moving hot
wind~\citep[but see][]{Zhang2017}, accelerated by radiation or cosmic ray
pressure~\citep{Murray2005, Kim2018}, or formed out of radiative cooling of the
hot wind material~\citep{Wang1995, Thompson2016}. Without ongoing star-forming
or quasar~\citep{Cai2019} activities, it is unclear whether galactic winds
remain a viable channel for producing cool CGM gas in massive quiescent
galaxies.

One possibility is that the galactic winds launched during the star formation
episodes prior to quenching may survive more than a few Gyrs. Recently,
\citet{Lochhaas2018} developed an 1D semi-analytic model of CGM as long-lived
bubbles blown by galactic winds, which then shock and sweep up ambient halo gas
as they propagate outward from the host galaxy~\citep[also see][]{Samui2008,
Sarkar2015}. They showed that the shocked wind can radiatively cool to $10^4$ K
and the cooled massive shells can travel to several hundred $\kpc$s with
$v{\sim}100{-}300\kms$ within a span of $1{-}10$ Gyrs. Therefore, the bubble
effectively ``hangs'' at a large distance from the galaxy for a long time,
explaining the presence of wind material in the CGM surrounding quiescent
galaxies.  
Since \citet{Lochhaas2018} mainly focused on the Milky Way-sized haloes with
$M_h{\simeq}10^{12}\msol$ probed by COS-HALO~\citep{Tumlinson2013}, it is
unclear if the results also applies to the LRG-size haloes with
$M_h{\gtrsim}10^{13}\msol$.

Apart from galactic winds, cool clouds could also appear in the CGM of massive
quiescent galaxies via cosmological gas accretion~\citep{Keres2009,
Fumagalli2011, vandeVoort2012}, gas stripping from satellite
infall~\citep{Gauthier2010, Gauthier2013, Lee2021}, and condensation due to
thermal instability in the hot medium~\citep{Mo1996, Maller2004, Sharma2012,
McCourt2012, Voit2015, Voit2018, Nelson2020}. All three scenarios above involve
the infall of cool clouds through the hot halo gas while experiencing effects of
hydrodynamic drag and cloud evaporation~\citep{Armillotta2017}, modulo
differences in the initial cloud conditions. For example, \citet{Afruni2019}
developed a semi-analytic model for cloud accretion from the intergalactic
medium, finding that the accretion of clouds with cloud mass
$\mcl{\sim}10^5\msol$ could reproduce the velocity distribution and column
densities of the cool clouds observed by COS-LRG~\citep{Chen2018, Zahedy2019}.
Meanwhile, those clouds will evaporate before reaching the host galaxy,
therefore unlikely rejuvenating any star formation. However, there is no
appreciable difference between the observed metallicity distributions of the
star--forming and quiescent galaxies~\citep{Lehner2013, Berg2019}. Therefore,
without the enrichment from stellar outflows, it is unclear how the newly
accreted cool gas becomes significantly metal-enriched while in pressure
equilibrium with the hot gas.

The key to distinguishing between the wind vs. no--wind origins lies in the
kinematics of the cool clouds. The most common probe of the cloud kinematics is
the distribution of redshifts of MgII absorbers relative to their host galaxies
$\Delta z$. In particular, $\Delta z$ is usually interpreted in the literature
as the relative line-of-sight~(LOS) velocity of MgII absorbers~(normalized by
the speed of light $c$), so that $c \Delta z \simeq \vlos (1+z) $. This
approximation, however, ignores the fact that $\Delta z$ also depends on the
comoving distance between the two systems along the LOS, $y$ , so that $c \Delta
z  = \vlos  (1+z)+ H_z y $\footnote{$y$ and $\vlos$ are both vectors, which
could be negative or positive depending on whether it is pointed toward or away
from the observer, respectively.}, where $H_z$ is the Hubble parameter at $z$.
The approximation is more than adequate for comparing the LOS velocity
dispersion of the MgII absorbers to that of the dark matter~\citep[see e.g.,
][]{Huang2016}, but for a more stringent test of cloud kinematic models we need
to explicitly model the redshift-space galaxy-absorber cross--correlation
functions~\citep{Lanzetta1998, Chen2005, RyanWeber2006, Wilman2007, Chen2009,
Tejos2012, Borthakur2016}.
In this paper, we will first reconstruct the 3D spatial distribution
of cool clouds from the observed projected LRG-MgII cross-correlation function,
and then infer the velocity distribution from the observed RSD of MgII absorbers
around LRGs. A similar method was first developed by ~\citet{Zu2013} for
inferring the galaxy infall kinematics around massive galaxy clusters.

This paper is organized as follows. We describe the BOSS LRG and the MgII
absorber catalogues in~\S\ref{sec:data}. The RSD measurements are presented
in~\S\ref{sec:rsd}. We present our reconstruction of the 3D distribution of MgII
absorbers within the LRG haloes and extract the small-scale RSD signal of MgII
absorbers in~\S\ref{sec:deproj}. We compare the theoretical predictions of RSD
from different kinematic models in~\S\ref{sec:kin} before we summarise our
results and look to the future in~\S\ref{sec:conc}.  Throughout this paper, we
assume the {\it Planck} cosmology~\citep{Planck2018}.  All the length and mass
units in this paper are scaled as if the Hubble constant is
$100\,\kms\mpc^{-1}$. In particular, all the separations are co-moving distances
in units of $\hmpc$, and the halo mass is in the unit of $\hmsol$. We use $\lg
x{=}\log_{10} x$ for the base-$10$ logarithm.

\section{Data}
\label{sec:data}

\begin{figure}
\begin{center}
    \includegraphics[width=0.48\textwidth]{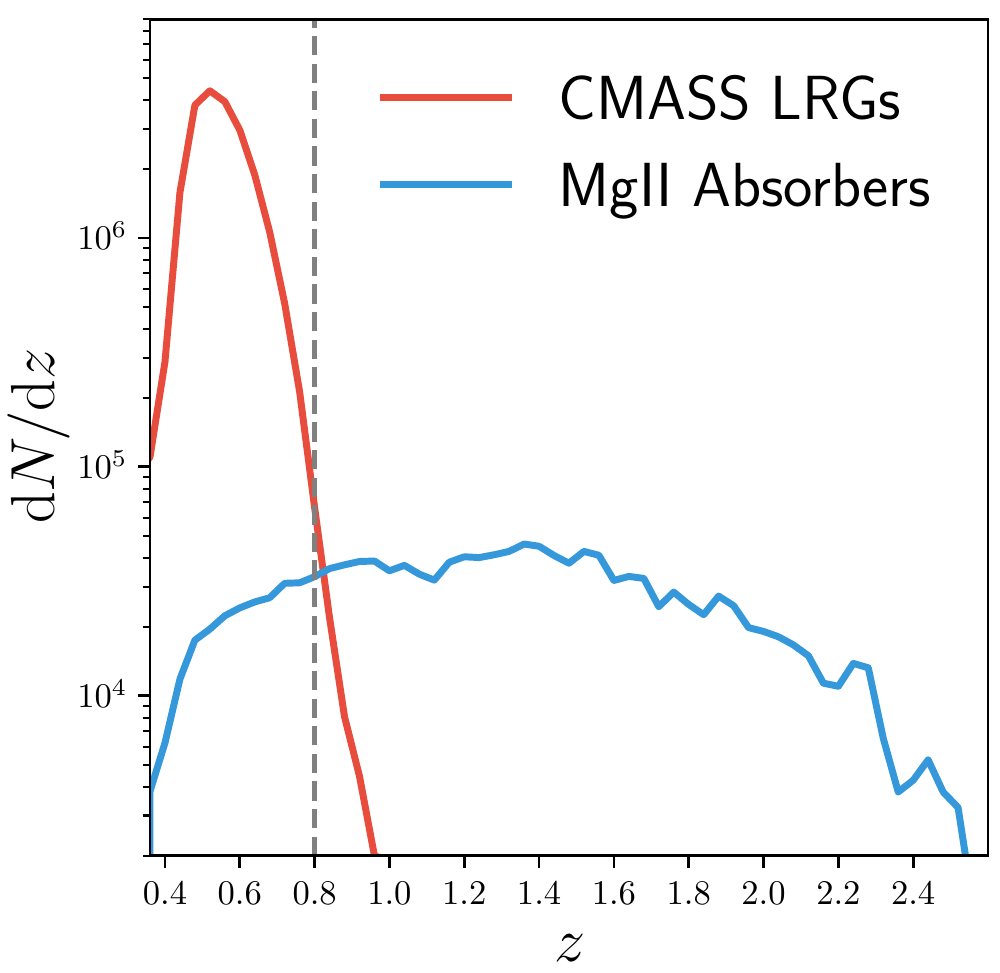} \caption{
The redshift distributions of CMASS LRGs~(red) and MgII
absorbers~(blue) between $z{=}0.4$ and $2.5$. The gray dashed vertical line indicates the
maximum redshift~($z{=}0.8$) of our cross-correlation analysis.}
\label{fig:nz}
\end{center}
\end{figure}

\subsection{Luminous Red Galaxies from BOSS CMASS}
\label{subsec:lrg}

We employ the BOSS CMASS galaxy sample derived from the Data Release
12~\citep[DR12;][]{Alam2015} of Sloan Digital Sky Survey~\citep[SDSS;][]{York2000}.  As
part of the SDSS-III~\citep{Eisenstein2011}, BOSS~\citep{Dawson2013} measured
the spectra of 1.5 million galaxies over a sky area of ${\sim}$10,000 deg$^{2}$
using the BOSS spectrographs~\citep{Smee2013, Bolton2012} onboard the 2.5-m
Sloan Foundation Telescope at the Apache Point Observatory~\citep{Gunn1998,
Gunn2006}. The BOSS LRGs were selected from the Data Release
8~\citep[DR8;][]{Aihara2011} of SDSS five-band imaging~\citep{Fukugita1996}
using two separate sets of colour and magnitude cuts for the
LOWZ~($z{<}0.4$) and CMASS~($z{>}0.4$) samples,
respectively~\citep{Reid2016}.  For maximal redshift overlap with the MgII
absorbers, we select $823193$ CMASS galaxies with $z\in[0.4, 0.8]$ as our LRG
sample for cross-correlating with the MgII absorbers. We generate a random
galaxy sample based on the standard CMASS random galaxy catalogue provided in
the public SDSS DR12 directory, by applying the redshift distribution of the
observed galaxy sample to the random catalogue~(red curve in
Figure~\ref{fig:nz}). To minimize the shot noise from the random, we use 50
times more galaxies in the random than in the observed data set.

\subsection{MgII Absorbers}
\label{subsec:mgii}

We select a sample of $9088$ MgII absorbers within the CMASS footprint with
$z\in[0.4, 0.8]$ from the JHU-SDSS MgII absorber
catalogue\footnote{https://www.guangtunbenzhu.com/jhu-sdss-metal-absorber-catalog}
derived by \citet{Zhu2013}, who applied an automatic absorption-line detection
algorithm to ${\sim}140000$ quasar spectra from the SDSS
DR5~\citep{Schneider2010} and DR12~\citep{Paris2017}. Using a novel nonnegative
matrix factorization technique to estimate the quasar continua, \citet{Zhu2013}
detected $61974$ unique MgII absorbers between $0.4<z<5.5$. The completeness and
purity of the absorber catalogue are further validated with the visually
inspected Pittsburgh MgII catalogue~\citep{Quider2011} derived from SDSS DR4.
For cross-correlating with the CMASS galaxies, we limit our sample to absorbers
within the CMASS footprint and with $z\in[0.4, 0.8]$, yielding $9088$ MgII
absorbers in the final sample. We do not select absorbers by their equivalent
width~(EW), but will apply the values of equivalent width as individual weights when
computing the correlation functions.

The MgII absorbers were detected along the sightlines of SDSS quasars, which
generally were given a higher target priority by the SDSS tiling
algorithm~\citep{Blanton2003} than galaxies.  Therefore, in order to build a
truly random catalogue of MgII absorbers, we need to mimic the spectroscopic
survey strategy of SDSS, so that the conditional fibre assignment probability of
quasars at any angular separation away from a foreground galaxy is exactly
preserved in the random absorber catalogue. Since this conditional probability
is analytically intractable, we make use of the parent quasar catalogue that
\citet{Zhu2013} searched for MgII absorbers, and randomly assign redshifts of
the detected MgII absorbers to the angular coordinates of the observed quasars.
To remove any potential systematics associated with the difference between DR7
and DR12 quasars, we assign the redshifts of DR7 absorbers to the coordinates of
DR7 quasars, and vice versa for DR12. This random generation scheme also
eliminates any systematic bias due to the intrinsic clustering of quasar
sightlines~\citep{Gauthier2010}. Finally, we shuffle the redshifts of the
detected MgII absorbers ten times to build a random absorber catalogue ten times
the size of the data catalogue.

Figure~\ref{fig:nz} shows the redshift distributions of CMASS LRGs~(red) and
MgII absorbers~(blue) between $z{=}0.4$ and $2.5$. The dashed vertical line
indicates the maximum redshift~($z{=}0.8$) of our analysis, beyond which the
number density of CMASS LRGs declines rapidly. We note that the number density
of MgII absorbers stays relative flat within the vast redshift range between
$z{=}0.8$ and $1.8$, leaving huge statistical potential to future surveys that
will target a high number density of LRGs and emission lines galaxies
within the same redshift range.

\section{Galaxy-cloud Cross-correlation Function in Redshift-space
$\mathbf{\xirs}$}
\label{sec:rsd}

\begin{figure}
\begin{center}
    \includegraphics[width=0.48\textwidth]{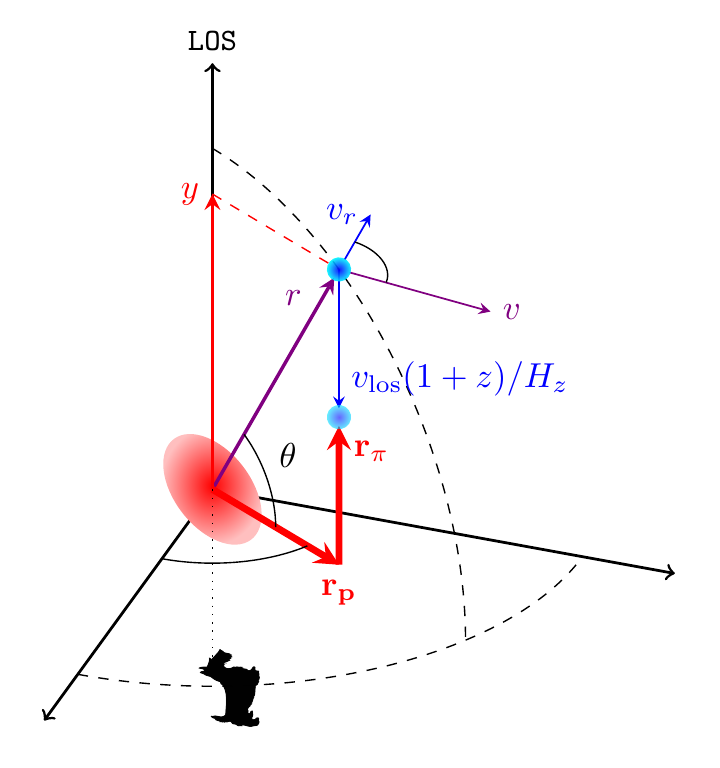}
    \caption{Schematic diagram of the redshift-space distortion of a MgII
    absorber~(dark-shaded blue circle) in the halo of an LRG~(red ellipse) at
    redshift $z$. The observer is at the bottom of the diagram, making the
    line-of-sight direction vertically upward. The LOS velocity $\vlos$ shifts
    the cloud from its real-space LOS distance $y$ by $\vlos (1+z)/ H_z$ to
    $r_{\pi}$~(light-shaded blue circle), where $H_z$ is the Hubble parameter at
    $z$.}
\label{fig:cloud_geometry}
\end{center}
\end{figure}

Before presenting the redshift-space cross-correlation function between CMASS
LRGs and MgII absorbers $\xirs$, we introduce the notations relevant to
describing cloud kinematics in the galacto-centric coordinates in
Figure~\ref{fig:cloud_geometry}.  The red ellipse indicates the position of the
LRG at redshift $z$, with the LOS direction pointing vertically upward. The
dark-shaded blue circle indicates the real-space position of a cool cloud, with
3D distance $r$ away from the LRG, which can be further decomposed into a
projected separation $r_p$ and an LOS distance $y$. However, unlike $r_p$, $y$
is not directly accessible through observations, because the relative velocity
of the cloud $v$ has an LOS component $\vlos$ that displaces the observed
position of the cloud~(light-shaded blue circle) to $r_\pi$ along the LOS by
$\vlos (1+z)/ H_z$. Obviously, to recover the kinematics of the cool
clouds~($\vlos$), we need to infer both the real and redshift--space
distributions of the cool clouds~($y$ and $r_{\pi}$).


\subsection{Correlation Function Measurements}
\label{subsec:measurement}

\begin{figure*}
\begin{center}
    \includegraphics[width=0.96\textwidth]{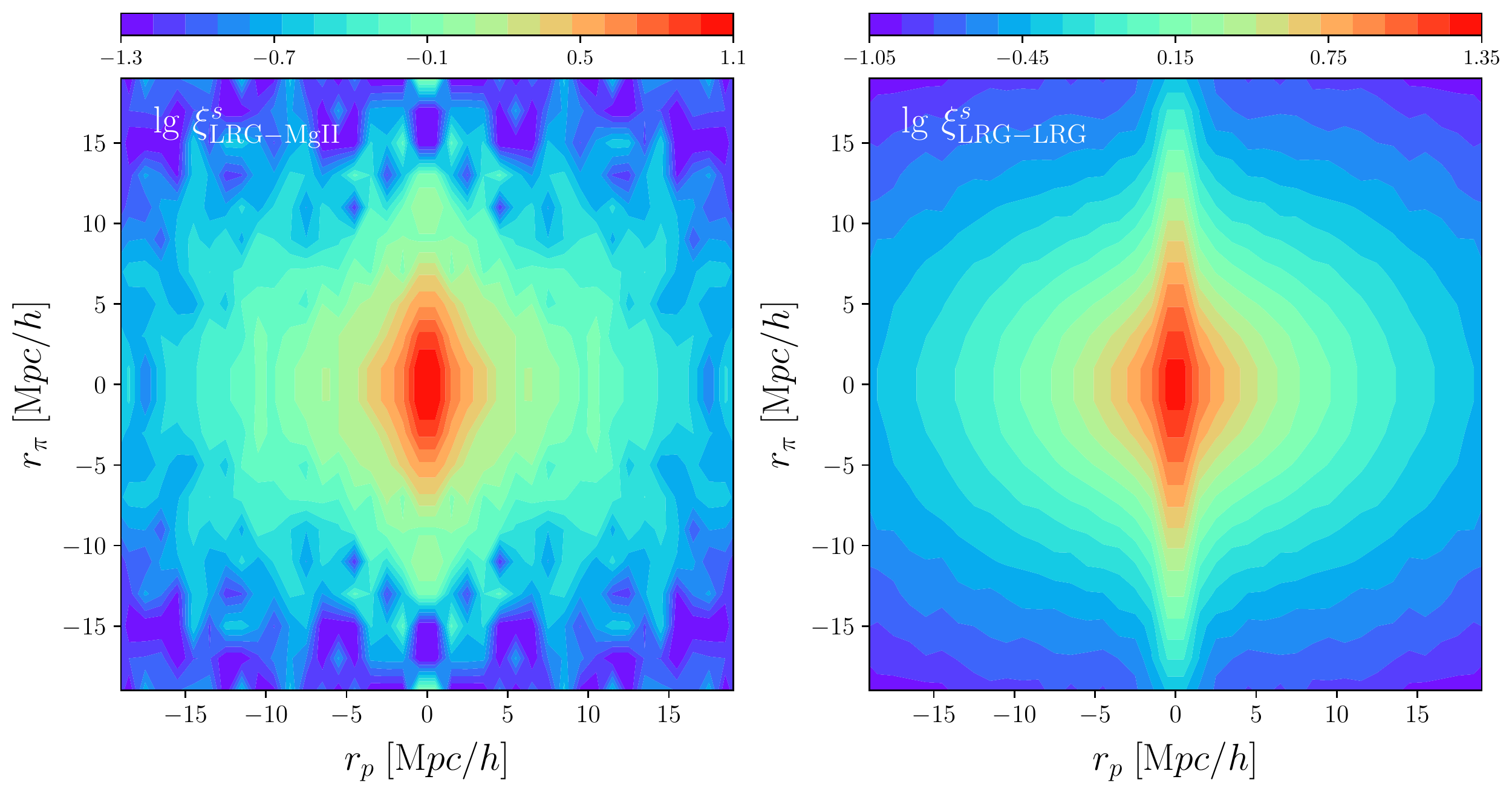}
    \caption{{\it Left}: Redshift-space cross-correlation function between LRGs
    and MgII absorbers; {\it Right}: Redshift-space auto-correlation
    function of LRGs. The logarithmic values of the correlation
    functions are indicated by the colour bars on top. The $0.25$ dex shift
    between the two colour bars accounts for the difference in the large-scale
    bias between MgII absorbers and LRGs. The small-scale Fingers-of-God
    effect is pronounced in the right panel, but weak in the left panel.}
\label{fig:rsd_comparison}
\end{center}
\end{figure*}

We measure $\xirs$ using the~\cite{Landy1993} estimator,
\begin{equation}
    \xirs(r_p,r_\pi) = \frac{D_g D_c - D_g R_c - R_g D_c + R_g R_c}{ R_g R_c} ,
    \label{eqn:ls}
\end{equation}
where $D_{g}D_{c}$ and $R_{g}R_{c}$ are the number counts of data galaxy--data
cloud pairs and random galaxy--random cloud pairs at 2D separation $(r_p,
r_\pi)$, respectively. By the same token, $D_{g}R_{c}$ and $R_{g}D_{c}$ indicate
the corresponding number counts of data galaxy-random cloud pairs and random
galaxy-data cloud pairs. For computing pair counts associated with the data clouds, we
weigh each pair by the EW of the MgII absorber~(the sum of the doublet EWs). We
also repeat our analysis in this paper using measurements without the EW weights, and
discover that the impact of weights on our conclusions is negligible.

To compare the RSD of MgII absorbers with that of galaxies, we also measure the
redshift--space galaxy auto--correlation functions via
\begin{equation}
    \xirsgg(r_p,r_\pi) = \frac{D_g D_g - 2 D_g R_g  + R_g R_g}{ R_g R_g}.
\end{equation}
The difference between $\xirs$ and $\xirsgg$ on large scales should be a
constant shift in amplitude by a factor of $b_g / b_c$, where $b_g$ and $b_c$
are the large-scale clustering biases of the LRGs and MgII absorbers,
respectively. On small scales, however, any difference between the two would be
primarily induced by the deviation of cloud kinematics from that of the luminous
red satellites within the haloes.

For reconstructing the 3D number density profile of the clouds, we measure the projected galaxy--cloud
cross--correlations $w_p$ from the same data and random galaxy samples using an
integration length of $r_\pi^\mathrm{max}=40\,\hmpc$, so that
\begin{equation}
    w_p(r_p) = \int_{-r_\pi^\mathrm{max}}^{+r_\pi^\mathrm{max}}\! \xirs(r_p,r_\pi)\,\dd r_\pi.
    \label{eqn:wp}
\end{equation}

Finally, we estimate the measurement uncertainties on $\xirs$ and $w_p$ using the
standard Jackknife re--sampling method by dividing the CMASS sky coverage into
$200$ approximately equal-area regions~\citep{Norberg2009}.

\subsection{Comparison between $\mathbf{\xirs}$ and $\mathbf{\xirsgg}$}
\label{subsec:rsdcomp}

Figure~\ref{fig:rsd_comparison} compares the RSD signature of the MgII
absorbers~(left) to that of the galaxies~(right) around LRGs, with the
logarithmic values of the correlation functions indicated by the respective
color bars on top. We shift the two color bars by $0.25$ index to account for
the bias difference between the two tracers, so that the Kaiser
effect~\citep{Kaiser1987} on scales beyond $r_p{\sim}5\,\hmpc$ looks similar
between the two panels, despite the much larger uncertainties in the measurement
of $\xirs$.

The most striking difference between the two 2D correlation functions is the
absence~(presence) of a strong Fingers-of-God~\citep[FOG;][]{Jackson1972} effect
in the RSD of MgII absorbers~(galaxies) around LRGs. This is probably not
surprising, given that we know the velocity dispersion of MgII absorbers is
smaller than that of dark matter within the LRG haloes~\citep{Zhu2014,
Huang2016, Anand2021, Huang2021}. However, it is unclear whether the lack of FOG
is caused simply by a reduced velocity dispersion, or some other kinematic
features that may allow us to distinguish between wind vs. no--wind origins of
the cool clouds.

\begin{figure*}
\begin{center}
    \includegraphics[width=0.96\textwidth]{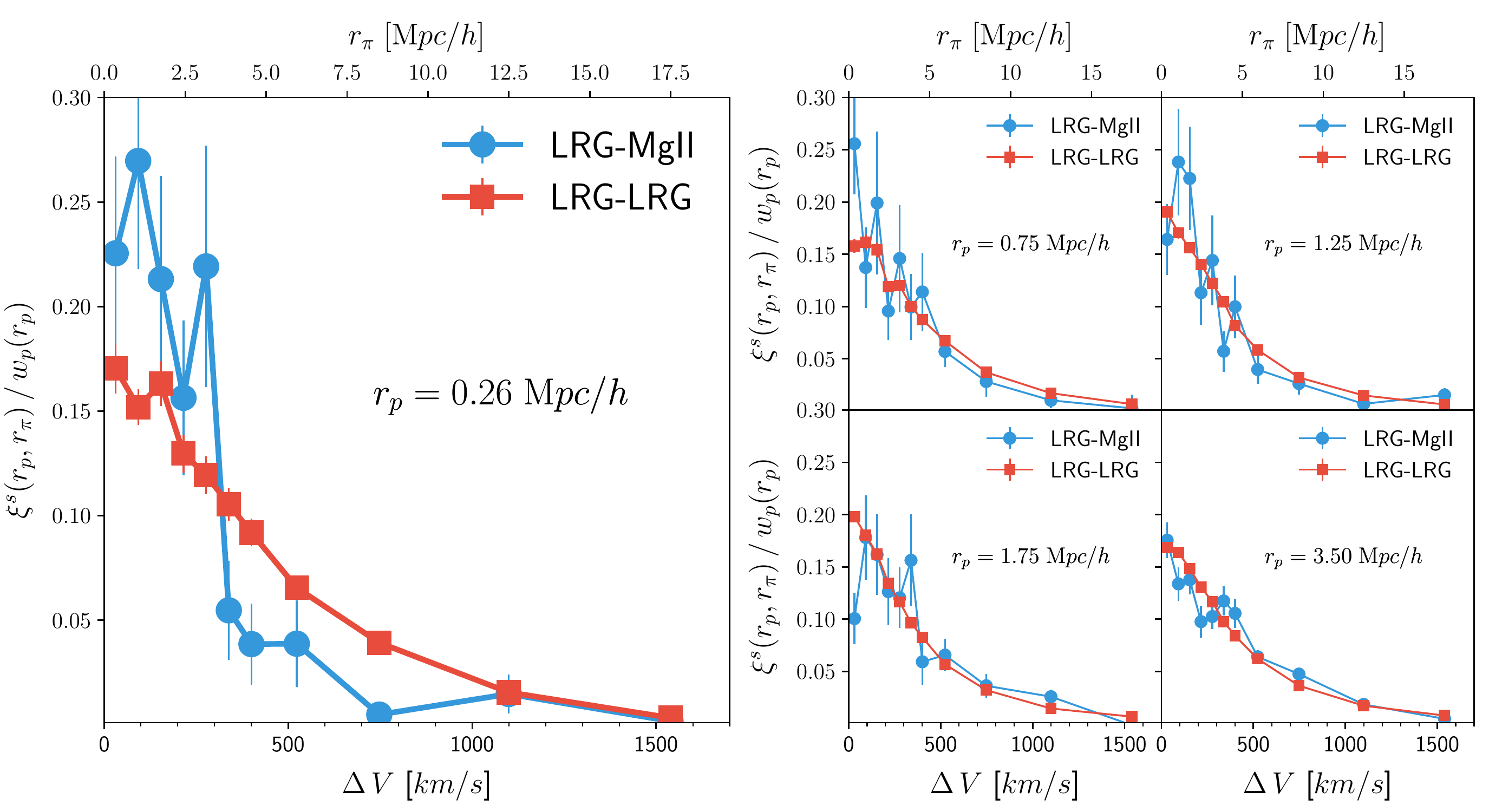}
    \caption{Comparison between the $\xi(r_p, r_{\pi}) / w_p(r_p)$ profiles of
    LRG--MgII cross--correlation~(blue circles) and LRG auto--correlation~(red
    squares), within the LRG haloes~(left main panel; $r_p{=}0.26\,\hmpc$) and
    on large scales beyond the 1--halo regime~(right panels; $r_p{=}[0.75, 1.25,
    1.75, 3.50]\,\hmpc$), respectively. For each panel, the bottom and upper
    x--axes indicate the LOS separation using the relative velocity $\Delta V$
    and the LOS distance $r_{\pi}$, respectively.  The two sets of profiles
    closely track each other at large $r_p$~(right panels), but differ
    significantly at $r_p{=}0.26$~(left panel), where a sharp truncation occurs
    at $\Delta V{\sim}300\,\kms$~(i.e., $r_{\pi}{\sim}0.34\,\hmpc$) in the
    LRG--MgII cross--correlation but not in the LRG auto--correlation.}
\label{fig:xirs_slice}
\end{center}
\end{figure*}

To look more closely into the small--scale differences in RSD, we calculate the
ratio between $\xi^{s}(r_p, r_{\pi})$ and $w_p(r_p)$ at fixed $r_p$, and compare the
two sets of ratios as functions of $r_{\pi}$ in Figure~\ref{fig:xirs_slice}. In
essence, the $\xi^{s}(r_p, r_{\pi})/w_p(r_p)$ ratio profiles measure the
redshift--space LOS probability distribution function of clouds or galaxies
around LRGs at any $r_p$. In the left panel of Figure~\ref{fig:xirs_slice}, we
compare the ratio profiles of the LRG--MgII cross--correlation~(blue circles)
and LRG auto--correlation~(red squares) at $r_p{\simeq}0.26\,\hmpc$~(i.e.,
$0.01\,\hmpc< r_p < 0.5\,\hmpc$). We limit our lowest $r_p$ bin below
$r_p{=}0.5\,\hmpc$ because we want to focus exclusively on the so--called
``1-halo'' term of the correlation signals --- the typical halo radius of the
LRG host haloes is estimated to be ${\sim}0.5{-}0.6\,\hmpc$ from either fitting
to the stacked MgII absorption profiles~\citep{Zhu2014} or the weak lensing of
massive quiescent galaxies~\citep{Mandelbaum2016, Zu2016}. We also compare the
two sets of ratio profiles computed at larger projected distances, $r_p{=}[0.75,
1.25, 1.75, 3.50]\,\hmpc$, in the four right panels. In each panel, we label the
bottom x--axis using the widely--used relative velocity $\Delta V$, while
showing the redshift--space LOS distance $r_{\pi}$ in the upper x--axis.

The four right panels of Figure~\ref{fig:xirs_slice} indicate that the LOS
distribution of MgII absorbers is in good agreement with that of galaxies at
$r_p{>}0.6\,\hmpc$, i.e., in the ``2--halo'' regime. This is unsurprising,
because the cool clouds should follow the distribution of galaxies on large
scales regardless of wind or no--wind origins. Within the LRG haloes~(left panel
of Figure~\ref{fig:xirs_slice}), however, the LOS distribution of MgII absorbers
is drastically different from that of galaxies, exhibiting a sharp truncation at
$\Delta V{\sim}300\,\kms$~(i.e., $r_{\pi}{\sim}3.4\,\hmpc$). This sharp
truncation explains the missing of FOG effect in the RSD of MgII absorbers, and
can not be reproduced simply by reducing the velocity dispersion of the cool
clouds~(as will be demonstrated later in~\S~\ref{sec:kin}).  We note that the
luminous red satellites do not exhibit the typical satellite kinematics within
the LRG haloes --- they are subject to stronger dynamic friction and likely had
an earlier infall, therefore travelling more slowly than the avearge satellites.
We thus expect the cross--correlation between the LRGs and typical galaxies
would produce an even stronger FOG effect than the LRG auto--correlation.  We
also note that \citet{Kauffmann2017} also measured the redshift--space
cross--correlation function between the CMASS LRGs and MgII absorbers. However,
they used a different method of generating random catalogues, a different
estimator than Equation~\ref{eqn:ls}, and a different binning, so it is
difficult to make direct comparison between Figure~\ref{fig:xirs_slice} and
their results.  To extract the kinematic information encoded in the $\xirs$
profile at $r_p{=}0.26\,\hmpc$, in the next step we will reconstruct the
real--space LOS distribution of MgII absorbers around the LRGs from the
measurements of $w_p(r_p)$.

\section{Spatial Distribution of Cool Clouds Around LRGs}
\label{sec:deproj}

\begin{figure*}
\begin{center}
    \includegraphics[width=0.96\textwidth]{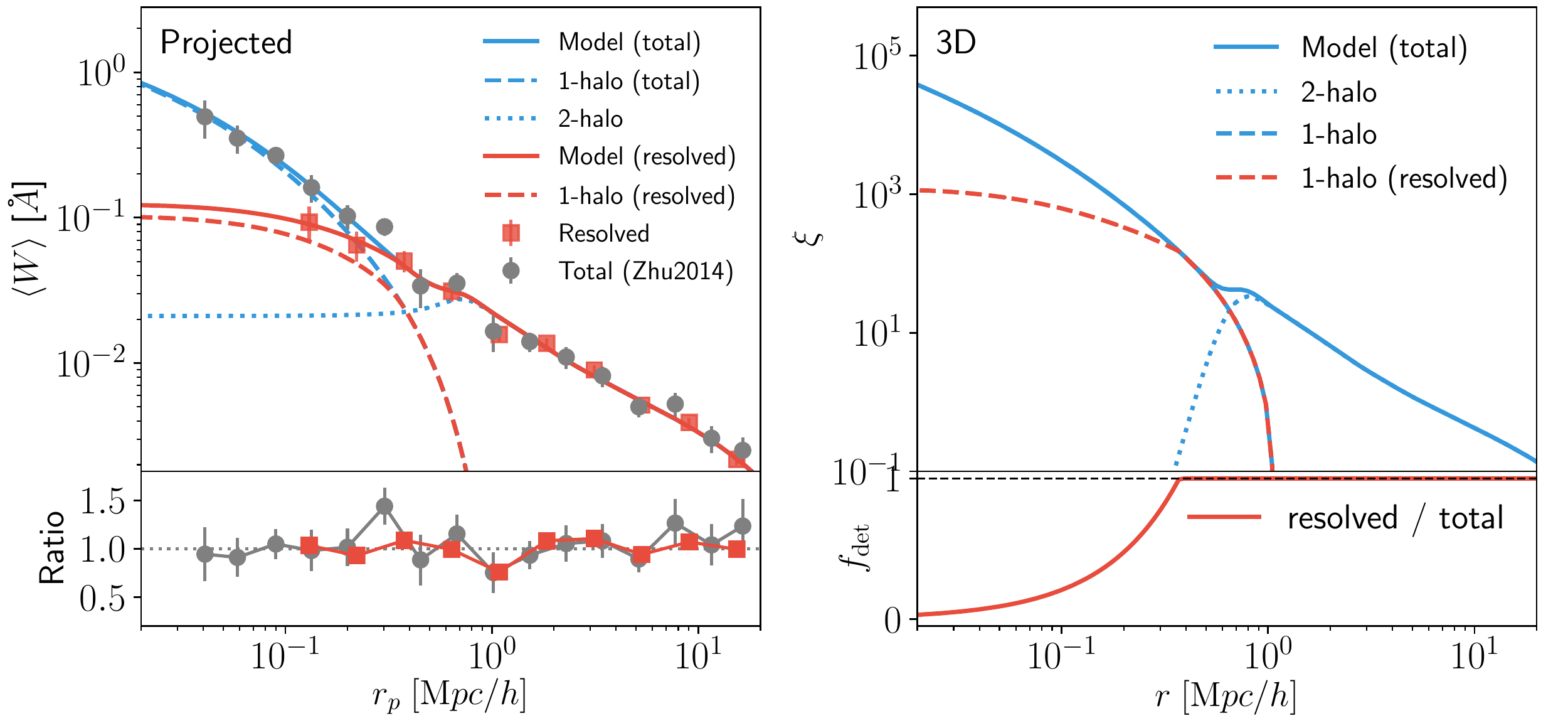} \caption{{\it Left
    panel}: Projected EW profiles of MgII absorbers around LRGs. Gray circles
    with errorbars are the measurements of the total absorption from stacked
    quasar spectra by \citet{Zhu2014}, while red squares with errorbars indicate
    our measurements from the projected cross--correlation function between LRGs
    and the resolved MgII absorbers. Solid curves are the best--fitting
    predictions for the total~(blue) and resolved~(red) absorption signals, with
    dashed and dotted curves indicating the 1--halo and 2--halo components,
    respectively. The ratio profiles between the data measurements and
    best-fitting predictions are shown in the bottom panel for the total~(gray
    circles) and resolved~(red squares) absorption signals, respectively.  {\it
    Right panel}: Similar to the left panel, but for the 3D isotropic
    cross--correlation functions between LRGs and MgII absorbers. The bottom panel
    shows the predicted radius dependence of absorber detection fraction in the
    \citet{Zhu2013} catalogue.}
\label{fig:wp}
\end{center}
\end{figure*}

\subsection{Predict Projected EW Profile $\mathbf{\avg{W}(r_p)}$}
\label{subsec:ew}

\citet{Zhu2014} measured the projected MgII EW profile around the same sample of
LRG galaxies, from the flux decrements induced in the stacked quasar spectra by
MgII absorbers at the redshifts of the foreground LRGs. The \citep{Zhu2014}
profile is subsequently confirmed by other stacked MgII absorption
measurements~\citep[also see][]{PerezRafols2015, Lan2018}, and is roughly
consistent with the stacked H$\alpha$ emission measurements, suggesting the
metal absorption and H$\alpha$ emission likely originate from the same CGM
gas~\citep{Zhang2016, Zhang2018}.  The stacked signal of \citet{Zhu2014} is
analogous to our measurement of the projected cross--correlation between LRGs
and MgII absorbers $w_p(r_p)$, but with one key difference.  That is, by
stacking the quasar spectra, \citet{Zhu2014} included the contribution from {\it
all} the MgII absorbers, while our $w_p$ measurement only includes the
spectrally {\it resolved} absorbers. Comparing to our method, the stacking
technique is more useful for characterizing the {\it total} gas mass density
profile, but at the expense of the individual redshifts of gas clouds, therefore
less useful for inferring cloud kinematics. To reconstruct the spatial
distribution of the resolved MgII absorbers, we will utilize the stacked
absorption signal to infer the total gas density profile, which can be separated
into contributions from the non-resolved vs. resolved MgII absorbers by fitting
to the $w_p$ measurement assuming a radius--dependent detection fraction of the
resolved absorbers.

To facilitate our joint analysis with the stacked MgII EW profile measured
by~\citet{Zhu2014}, we convert our $w_p(r_p)$ measurement into a projected MgII
EW profile $\avg{W}$,
\begin{equation}
    \avg{W}(r_p) = w_p(r_p) \frac{\dd W}{\dd r_{\pi}},
    \label{eqn:wp2w}
\end{equation}
where $\dd W/\dd r_{\pi}$ is the average EW of MgII absorption per $\hmpc$
along the LOS. Ideally, this quantity should be computed by summing the EWs of
all the MgII absorbers within the redshift range probed. However, the
\citet{Zhu2013} MgII absorber catalogue suffers severe incompleteness at
the low EW end, and an extra redshift--dependent incompleteness due to strong
sky lines and sensitivity variation of the SDSS spectrographs. Therefore, we empirically infer the value of $\dd
W/\dd r_{\pi}{=}2.6\times 10^{-4}\,h\angstrom / \mpc$, from the ratio between
the stacked MgII EW profile measured by~\citet{Zhu2014} and our measurement of
$w_p$ on scales beyond $3\,\hmpc$.

The left panel of Figure~\ref{fig:wp} compares the $\avg{W}$ profiles inferred
from $w_p$ via Equation~\ref{eqn:wp2w}~(red squares) and from the stacking
technique of \citet{Zhu2014}~(gray circles). The two measurements are roughly
consistent on scales above $0.5\,\hmpc$ --- this is by design because we have
adjusted the value of $\dd W/\dd r_{\pi}$ to match the two. However, the two
profiles start to diverge below $0.5\,\hmpc$. In particular, the $w_p$--inferred
$\avg{W}$ of resolved absorbers exhibits a flat trend with decreasing $r_p$,
while the stacked $\avg{W}$ of total absorption is consistent with a projected
NFW~\citep[Navarro-Frenk-White;][]{Navarro1997} profile~\citep{Zhu2014}.
The discrepancy between the two $\avg{W}$ profiles on small scales implies that there is
a radius-dependent detection inefficiency of the Zhu \& Menard catalogue, so that MgII
absorbers that are closer to the LRGs are less likely to be detected. At fixed EW, the close
proximity of quasar sightlines to LRGs shoud not affect the detection probability of MgII
absorbers, because the scattered light from the LRGs into the quasar fibres is minimal. We
speculate that the radius-dependent inefficiency is induced by the fact that the average
EW of the MgII absorbers decreases from the outer CGM to the center of the LRGs. Since the
detection probability decreases rapidly at $\mathrm{EW} < 2 \mbox{\normalfont\AA}$~(see
figure 7 of \citet{Zhu2013}), the stacked MgII absorption profile is much steeper than the
number density profile of the resolved MgII absorbers.
The overall shape
of our $w_p$--inferred $\avg{W}$ profile is consistent with the excess absorber
number density profile measured by \citet{Anand2021}, who utilized a larger MgII
absorber catalogue derived from the SDSS DR16.

For the purpose of our analysis, it is more convenient to work with correlation
functions than density profiles. Therefore, instead of reconstructing the radial
number density profile of MgII absorbers $n(r)$, we focus on the ``1--halo''
component of the 3D isotropic cross--correlation function between the LRGs and
absorbers $\xi_{1h}(r)$. The two are related via
\begin{equation}
    \xi_{1h}(r) = \frac{n(r) - \overline{n}}{\overline{n}},
\end{equation}
where $\overline{n}$ is the mean number density of cool clouds.

Following \citet{Zhu2014}, we assume that the {\it total} cool gas distribution follows
the dark matter distribution within haloes, which we model as
\begin{equation}
    \xi_{1h}^{\mathrm{gas}}(r) \equiv \xi_{1h}^{\mathrm{dm}}(r) = \frac{\rho_{\mathrm{NFW}}(r \mid M_h, c)}{\rho_{m}}f_{\mathrm{trans}}(r) - 1,
    \label{eqn:xi1hgas}
\end{equation}
where $\rho_{\mathrm{NFW}}(r\mid M_h,c)$ is the NFW density profile of a dark matter
halo with halo mass $M_h$ and concentration $c$, $\rho_m$ is the mean density of the
Universe, and $f_{\mathrm{trans}}$ describes the so--called ``spashback'' transition near the halo boundary~\citep{Diemer2014},
\begin{equation}
    f_{\mathrm{trans}}(r) = \left[ 1 + \left(\frac{r} {1.495\, r_{200m}}\right)^4\right]^{-2}.
\end{equation}
We emphasize that although the result from \citet{Zhu2014} shows that an NFW
profile provides an adequate description of the average cool gas density
profile, the spatial distribution of the cool gas clouds in {\it individual}
haloes could deviate significantly from the dark matter, depending on the
dynamical state of the cool gas. For example, \citet{Wild2008} discovered that
quasars could destroy MgII absorbers up to comoving distances of ${~}800$ $kpc$
along their lines of sight, so that the spatial distribution of cool clouds
could depend strongly on the duty cycle of the supermassive blackholes in the
galactic centre.  However, it is plausible that the gas density profiles
averaged over many haloes of various dynamical states may converge to an
NFW--like profile. This possibility can be tested in the future against the
increasingly fine-grained hydrodynamic simulations of the CGM~\citep{Peeples2019,
Hafen2019, Suresh2019, Hummels2019, Fielding2020}.

For describing the {\it resolved} MgII absorbers in our sample, we further include a
radius--dependent inefficiency
\begin{equation}
    \xi_{1h}(r) = \frac{\rho_{\mathrm{NFW}}(r \mid M_h, c)}{\rho_{m}}f_{\mathrm{trans}}(r)f_{\mathrm{det}}(r) - 1,
    \label{eqn:xi1h}
\end{equation}
where $f_{\mathrm{det}}$ is the detection fraction
\begin{equation}
    f_{\mathrm{det}}(r) = \left(\frac{r}{\kappa\, r_{200m}}\right)^{\alpha},
\end{equation}
and we set $f_{\mathrm{det}}$ to be unity when $r{>}\kappa\, r_{200m}$.

We model the ``2--halo'' component of the 3D isotropic LRG--MgII cross--correlation as
\begin{equation}
    \xi_{2h}(r) = b_g b_c \xi_{mm}(r) f_{\mathrm{ex}}(r),
    \label{eqn:xi2h}
\end{equation}
where $\xi_{mm}$ is the dark matter auto--correlation function.  Since we do not
need to distinguish between $b_g$ and $b_c$, we merge them to form a single
parameter $b{\equiv}b_g b_c$. The last term on the right-hand side,
$f_{\mathrm{ex}}(r)$, describes the so--called ``halo exclusion'' effect, which
arises from the fact that the centre of one halo cannot sit within the virial
radius of another. In particular, $f_{\mathrm{ex}}(r)$ is unity at
$r{>}r^{\mathrm{max}}_{200m}$, as there is no halo exclusion when the distance
is larger than twice the radius of the largest haloes $r^{\mathrm{max}}_{200m}$.
By the same token, $f_{\mathrm{ex}}(r)$ declines to zero at some characteristic
distance that corresponds to twice the radius of the smallest haloes.

For analytically solving for $f_{\mathrm{ex}}$ at any given distance, we need to
explicitly integrate over all pairs of haloes that are too small
to run into each other when separated by that distance\citep[e.g., see equation
B10 of][]{Tinker2005}. However, such analytic calculation can only be performed if the
halo occupation distributions~(HODs) of both the LRGs and MgII absorbers are assumed {\it a priori}.
Since a comprehensive HOD modelling of the MgII absorbers is beyond the scope of this paper,
for our first--cut analysis we choose to use a simple error function to mimic the
behavior of halo exclusion
\begin{equation}
    f_{\mathrm{ex}}(r) = \frac{1}{2} \left[1 + \mathrm{erf}\left(\frac{r -
    \beta\,
    r_{200m}}{\sigma_{\mathrm{ex}}}\right) \right],
    \label{eqn:fex}
\end{equation}
where $\beta\,r_{200m}$ is the characteristic scale above which
$f_{\mathrm{ex}}$ asymptotes to unity, and $\sigma_{\mathrm{ex}}$ describes the
width of the transition of $f_{\mathrm{ex}}$ from unity to zero, which depends
on the widths of the halo mass distributions of the LRGs and MgII absorbers. We
have verified that for the HOD of a typical stellar--mass thresholded sample
in~\citep{Zu2015}, the analytic halo exclusion profile can be reasonably well
described by Equation~\ref{eqn:fex}.

We derive the overall $\xi(r)$ signals by directly summing the two components, so that
\begin{equation}
    \xi^{\mathrm{gas}}(r) = \xi_{1h}^{\mathrm{gas}}(r) + \xi_{2h}(r),
\end{equation}
for the LRG--gas cross--correlation,
and
\begin{equation}
    \xi(r) = \xi_{1h}(r) + \xi_{2h}(r),
\end{equation}
for the LRG--MgII absorbers cross--correlation, respectively. Note that we adopt
the same ``2--halo'' component for the total and resolved absorption, assuming
that the EW--dependence of cloud bias $b_c$ is
weak~\citep{Tinker2008, Gauthier2009}.  Finally, we integrate
the $\xi$ profiles along the LOS to obtain $w_p(r_p)$ via Equation~\ref{eqn:wp},
and then multiply by $\dd W/\dd r_{\pi}$ to predict the $\avg{W}(r_p)$ profiles.

\subsection{Reconstruct 3D Galaxy--cloud Cross--correlation}
\label{subsec:xi1h}

In total, we have seven free parameters, including $\{M_h, c, b, \kappa, \alpha,
\beta, \sigma_{\mathrm{ex}}\}$.  We perform simple $\chi^2$--minimization
fitting to the observed total and resolved $\avg{W}$ profiles, using the
diagonal errors provided by \citet{Zhu2014} for the stacked signal and the full
covariance matrix measured from Jackknife re-sampling for the resolved profile,
respectively. We enforce the 2--halo term to be zero at $r{<}0.2\hmpc$ during
the fit, in order to eliminate the false solution that artificially boosts the
2--halo term on very small scales to fit the data.  We find the best--fitting
values to be $\{\lg\,M_h{=}13.19, c{=}5.1, b{=}1.15, \kappa{=}0.63,
\alpha{=}1.2, \beta{=}1.15, \sigma_{\mathrm{ex}}{=}0.14\}$, and the
best--fitting predictions of the total~(blue solid) and resolved~(red solid)
signals are shown in the left main panel of Figure~\ref{fig:wp}. The bias value
is slightly higher than predicted for haloes with $\lg\,M_h{=}13.19$, probably
because a good fraction of the LRGs are satellite galaxies inside massive
haloes~\citep{Zu2020}. Both predictions are in excellent agreement with the
measurements, as demonstrated by the ratios between data and predictions in the
left bottom panel of Figure~\ref{fig:wp}.

In many observations, MgII absorbers detected at small impact
parameters~($r_p{<}0.2\,\hmpc$) to a foreground galaxy are usually considered
part of the CGM of that galaxy, but some of them should be interlopers from
other galaxies, i.e., part of the ``2--halo'' component. To obtain an estimate
of the interloper fraction, we show the best-fitting 1--halo components of the
total~(blue dashed) and resolved~(red dashed) absorption in the left panel of
Figure~\ref{fig:wp}. For the stacked measurement, the strong absorption signal
at $r_p{<}0.2\,\hmpc$ is almost entirely contributed by the gas clouds within
the LRG haloes~(i.e., the blue dashed curve is much higher than the blue dotted
curve).  For the resolved absorbers, however, ${\sim}1/3{-}1/5$ of the strong MgII
absorbers with impact parameters~(${<}0.2\,\hmpc$) are interlopers from the CGM
of other galaxies~(compare the red dashed curve with the blue dotted curve). This is roughly consistent with the expectation from
hydrodynamic simulations, where \citet{Ho2020} found that a ${\pm}500\,\kms$ cut
by $\vlos$ could include many gas clouds physically outside of the virial
radius.  Fortunately, our cross--correlation method is able to statistically
remove the impact of those interloper on our analysis.

In the right panel of Figure~\ref{fig:wp}, we show the best--fitting 3D
isotropic cross--correlation functions for the total signal~(blue solid), which
is further decomposed into 1--halo~(blue dashed; Equation~\ref{eqn:xi1hgas}) and
2--halo~(blue dotted; Equation~\ref{eqn:xi2h}) components; The 1--halo component
of the best--fitting $\xi$ profile of the resolved clouds is indicated by the
red dashed curve~(Equation~\ref{eqn:xi1h}).  In the bottom sub-panel, we show
the best--fitting profile of $f_{\mathrm{det}}$, indicating that the relatively
flat trend of $\avg{W}(r_p)$ below $r_p{\sim}0.3\,\hmpc$ is due to the
cancellation of the steep increase in total gas density by the rapid decline of
detection sensitivity toward small radii.

\subsection{Subtract Two--halo Term from $\mathbf{\xirs}$ }
\label{subsec:subtract}

\begin{figure}
\begin{center}
    \includegraphics[width=0.48\textwidth]{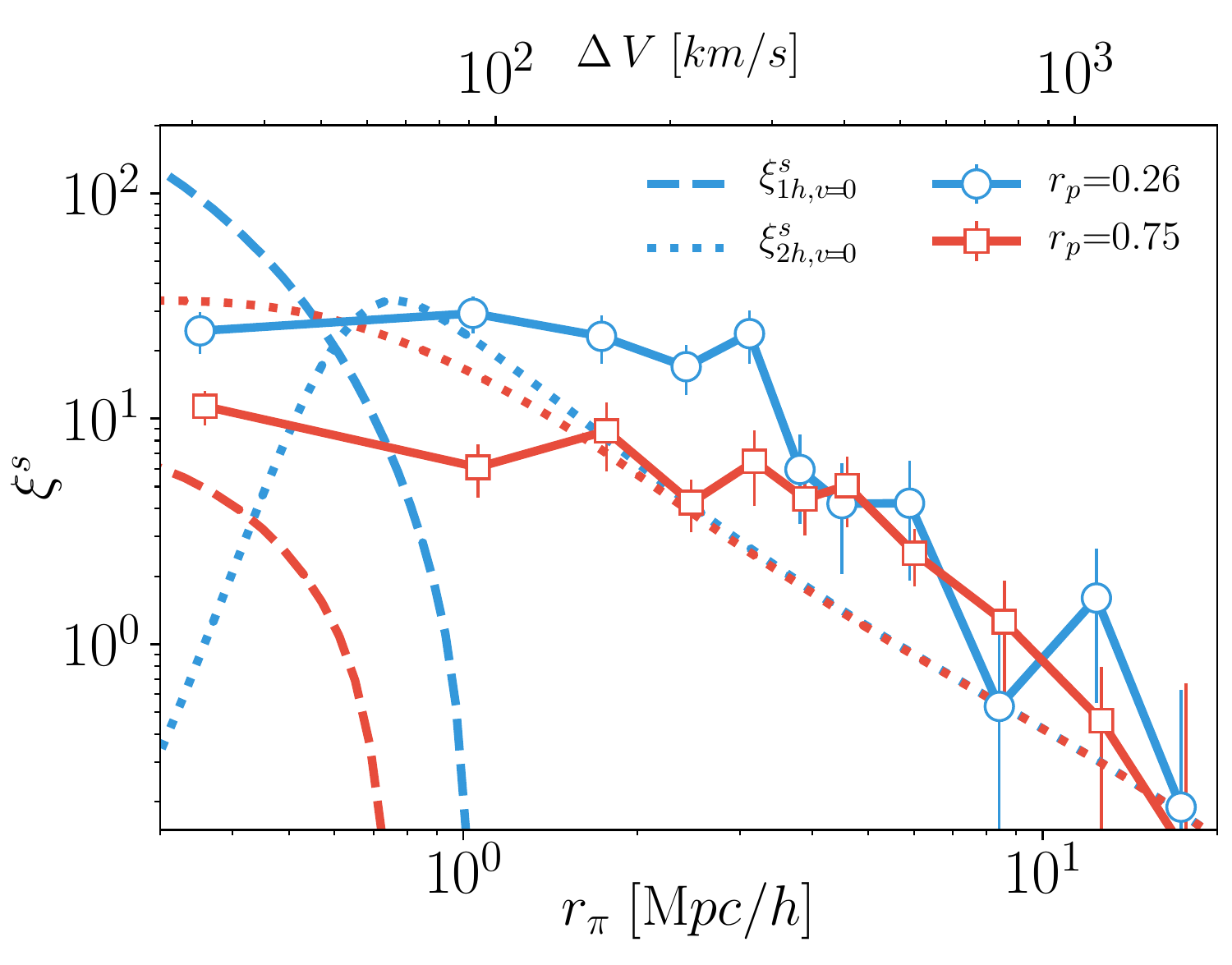}
    \caption{Comparison between the redshift--space LRG--MgII cross--correlation
    functions at $r_p{=}0.26\,\hmpc$~(blue) vs. $r_p{=}0.75\,\hmpc$~(red).
    Circles~(squares) with errorbars are the observed $\xi^{s}$ at
    $r_p{=}0.26$~($r_p{=}0.75$). Dashed and dotted curves indicate the 1--halo
    and 2--halo components of $\xi^{s}(r_{\pi}|r_p)$, respectively, predicted from
    $\xi(r{=}\sqrt{r_p^2{+}r_{\pi}^2})$, i.e., assuming zero peculiar velocities. We
    expect the 2--halo component of the observed $\xi^{s}$ signal at $r_p{=}0.26$
    can be well approximated by the $\xi^{s}$ signal observed at $r_p{=}0.75$.}
\label{fig:novpec}
\end{center}
\end{figure}

The kinematics of cool clouds within the LRG haloes is encoded in the
redshift--space cross--correlation function between the LRGs and MgII absorbers
$\xirs$ at small $r_p$. From the modelling of $w_p$ we infer that the typical
radius of the LRG haloes is ${\sim}0.6\,\hmpc$~(for haloes with
$\lg\,M_h{=}13.19$). Therefore, we will focus on the average RSD signal measured between
$r_p{=}0.01\,\hmpc$ and $0.5\,\hmpc$, i.e., $\xi^{s}(r_{\pi}|r_p{=}0.26\,\hmpc)$.

Ideally, one would model the full RSD signal on both small and large $r_{\pi}$,
i.e., including both the 1--halo and 2--halo components of $\xirs$ at
$r_p{=}0.26\,\hmpc$, as was previously done in, e.g., \citet{Zu2013}. However,
since we are only concerned with the CGM kinematics {\it inside} the haloes, we
will isolate the 1--halo term of the $\xirs$ signal by subtracting the 2--halo
term from the full RSD signal observed at $r_p{=}0.26\,\hmpc$. The isolated
1--halo RSD can then be modelled as the convolution between $\xi_{1h}$, the
quantity we have reconstructed in ~\S~\ref{subsec:xi1h}, and the LOS velocity
distribution of the cool clouds, the quantity we want to infer in this paper.

\begin{figure}
\begin{center}
    \includegraphics[width=0.48\textwidth]{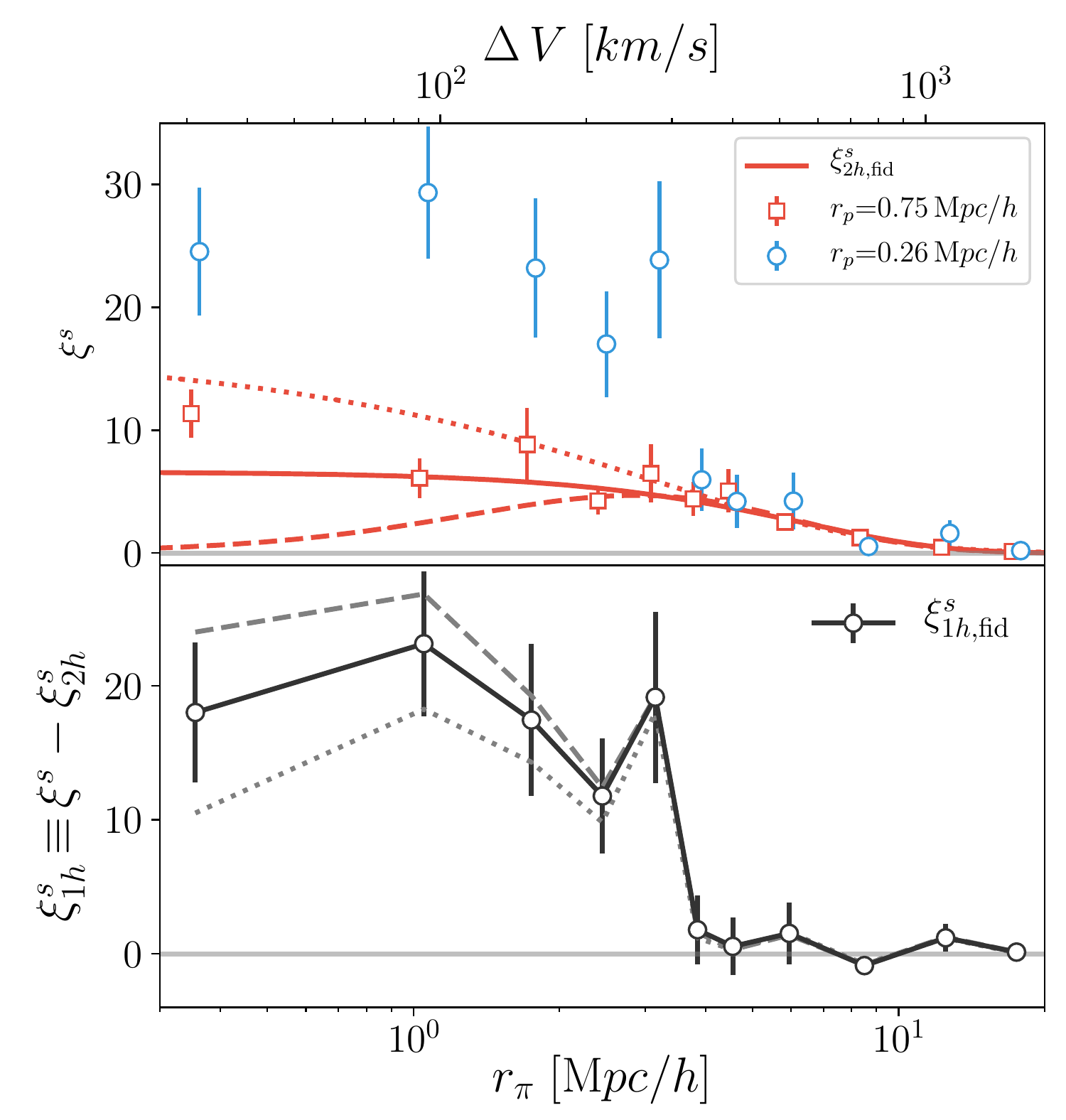} \caption{{\it
    Top panel}: Blue circles and red squares are the redshift--space LRG--MgII
    cross--correlation functions measured at $r_p{=}0.26\,\hmpc$ and
    $0.75\,\hmpc$, respectively. Red solid curve is our fiducial prediction of
    $\xi^{s}_{2h}$ from the best--fitting powered exponential
    function~(Equation~\ref{eqn:pe}). Red dotted and dashed curves are the upper
    and lower bounds of $\xi^{s}_{2h}$, respectively, inferred from adjusting
    the small--scale slope parameter of Equation~\ref{eqn:schetcher}. {\it
    Bottom panel}: The 1--halo components of
    $\xi^{s}(r_{\pi}|r_p{=}0.26\,\hmpc)$, obtained by subtracting the predicted
    $\xi^{s}_{2h}$ curves shown in the top panel from the observed signal at
    $r_p{=}0.26\,\hmpc$. Black circles with errorbars indicate our fiducial estimate of
    the $\xi^{s}_{1h}(r_{\pi}|r_p{=}0.26\,\hmpc)$ profile.}
\label{fig:subtraction}
\end{center}
\end{figure}

Accurate subtraction of the 2--halo term requires theoretically modelling the
large-scale infall of cool clouds outside the halo radius, which is highly
challenging and therefore beyond the scope of this paper.
For a first--cut
analysis, we derived the 2--halo term $\xi^{s}_{2h}(r_{\pi}|r_p{=}0.26\,\hmpc)$
empirically, by making use of the RSD signal observed at $r_p{=}0.75\,\hmpc$.
The rationale is
as follows. Firstly, without the peculiar velocities, the cross--correlation
function along the LOS is simply $\xi^{s}_{v{=}0}(y | r_p) =
\xi(\sqrt{y^2{+}r_p^2})$.  Since $\xi(r)$ is a slow--varying function of $r$ on
large scales, the underlying cloud distribution along the LOS should be similar
between $r_p{=}0.26$ and $0.75$ for $y{\gtrsim}2\,\hmpc$. Secondly, the LOS
velocity distribution is the projection of the isotropic 2D velocity
distribution along the LOS.  The two average ``sightlines'' at $r_p{=}0.26\,\hmpc$ and
$0.75\,\hmpc$ are almost parallel at the mean redshift of the LRG
sample~($z{\sim}0.55$), differing only by 1 arc min. Therefore, the LOS velocity
distributions of clouds beyond the halo radius should be almost the same between
the two $r_p$ bins. Combining the two factors together, we expect that the
2--halo components of the redshift--space cross--correlation signals are similar
between the two adjacent $r_p$ bins at $0.26\,\hmpc$ and $0.75\,\hmpc$.

Figure~\ref{fig:novpec} illustrates the rationale behind our expectation that
the two adjacent $r_p$ bins should share similar 2--halo terms of the RSD signal
around LRGs. The dashed and dotted curves indicate the 1--halo and 2--halo
contributions to $\xi^{s}_{v{=}0}(r_{\pi} | r_p)$, predicted by the
best--fitting $\xi(r)$ in the absence of peculiar velocities, at
$r_p{=}0.26\,\hmpc$~(blue) and $0.75\,\hmpc$~(red), respectively. Circles with
errorbars are the measured redshift--space correlation functions $\xirs$ for the
two $r_p$ bins~(also shown in Figure~\ref{fig:xirs_slice}). Unsurprisingly, the
1--halo term~(dashed blue curve) at $r_p{=}0.26\,\hmpc$ is ${\sim}20$ times
higher than that at $r_p{=}0.75\,\hmpc$~(dashed red curve), driving the strong
tophat--like enhancement of the observed redshift--space correlation on scales
below $r_{\pi}{\simeq}4\,\hmpc$~(i.e., $\Delta V{\simeq}300\,\kms$) for the
$r_p{=}0.26\,\hmpc$ bin. Furthermore, the 2--halo terms of the two $r_p$ bins
are indeed comparable when peculiar velocities are turned off~(dotted curves).
More important, the observed redshift--space correlations of the two $r_p$ bins
are highly consistent with each other at $r_{\pi}{>}4\,\hmpc$. This consistency
is very encouraging --- the underlying 2--halo component of the
observed $\xirs$ signal at $r_p{=}0.26$~(blue circles) should be well approximated
by the total observed $\xirs$ signal at $r_p{=}0.75$~(red squares).

Therefore, in order to estimate the underlying 1--halo term at
$r_p{=}0.26\,\hmpc$, we fit an analytic curve to the total observed $\xirs$
signal at $r_p{=}0.75\,\hmpc$ and subtract it from the total observed signal at
$r_p{=}0.26\,\hmpc$.  \citet{Zu2013} discovered that the $\xi^{s}(r_{\pi}|r_p)$
profiles can be well described by a powered exponential function
\begin{equation}
    \xi^{s}(r_{\pi}|r_p) = \xi^{s}_0 \exp \left\{-\left|\frac{r_\pi}{\rpic}\right|^\beta\right\},
    \label{eqn:pe}
\end{equation}
where $\rpic$ is the characteristic $r_{\pi}$ at which $\xi^{s}$ declines to
$1/e$ of its maximum value $\xi^{s}_0$ at $r_{\pi}{=}0$, and $\beta$ is the
shape parameter that controls the rate of the decline. We apply Equation~\ref{eqn:pe} to
the RSD signal at $r_p{=}0.75\,\hmpc$, and the results are shown in
Figure~\ref{fig:subtraction}.

In the top panel of Figure~\ref{fig:subtraction}, blue circles~(red squares)
with errorbars are the observed total $\xirs$ signals at
$r_p{=}0.26\,(0.75)\,\hmpc$. Red solid curve is the prediction from
Equation~\ref{eqn:pe} using the best--fitting parameters of $\rpic{=}6.71$,
$\beta{=}1.36$, and $\xi^{s}_0{=}41.67$. We omit the data point from the lowest
$r_{\pi}$ bin during the fit, where we expect a relatively larger discrepancy
between the 2--halo term at $r_p{=}0.26\,\hmpc$ and the observed signal at
$r_p{=}0.75\,\hmpc$ than other $r_{\pi}$ bins~(see Figure~\ref{fig:novpec}).
We also show two other model curves~(dotted and dashed) in which we use a different function form
\begin{equation}
    \xi^{s}(r_{\pi}|r_p) = \xi^{s}_0 \left|\frac{r_\pi}{\rpic}\right|^\alpha \exp \left\{-\left|\frac{r_\pi}{\rpic}\right|^\beta\right\},
    \label{eqn:schetcher}
\end{equation}
where the extra parameter $\alpha$ modifies the slope at small $r_{\pi}$. We
adjust the two model fits so that the dashed and dotted curves roughly delineate
the lower and upper bounds of the observed data points at $r_{\pi}{<}5\,\hmpc$,
respectively.

We show the derived 1--halo components of the $\xirs$ at $r_p{=}0.26\,\hmpc$ in
the bottom panel of Figure~\ref{fig:subtraction}, by subtracting each of the
three model curves of $\xi_{2h}^{s}$ shown in the top panel from the observed
$\xirs$ data points. Circles with errorbars represent our fiducial estimates for
$\xi^{s}_{1h}(r_{\pi} | r_p{=}0.26)$, obtained from the subtraction of the best
fit using Equation~\ref{eqn:pe}, while the dashed and dotted curves from the
subtraction of the respective curves predicted by Equation~\ref{eqn:schetcher}.
We emphasize that, despite the large discrepancies in $\alpha$, all three
estimates of $\xi^{s}_{1h}$ exhibit a sharp truncation at
$r_{\pi}{\sim}3{-}4\,\hmpc$, echoing the missing--FOG effect seen in
Figure~\ref{fig:rsd_comparison}. This also suggests that the sharp truncation is
physical, and most likely driven by some kinematic feature in the velocity
distribution of cool clouds. In the next section, we will describe the fiducial
measurements of $\xi^{s}_{1h}(r_{\pi} | r_p{=}0.26)$~(black open circles with
errorbars in the bottom panel of Figure~\ref{fig:subtraction}) using four
different cloud kinematic models. For our first--cut analysis, we directly apply
the Jackknife uncertainties of $\xi^{s}(r_{\pi}|r_p{=}0.26\,\hmpc)$ for the
error matrix of the fiducial measurements, i.e., assuming zero uncertainty from
the subtraction of the 2--halo term. Despite that the errorbars are
underestimated, they are roughly consistent with the allowed range sandwiched
between the lower and upper bounds, indicated by the gray dotted and dashed
curves, respectively.

\section{Comparison of Cloud Kinematic Models}
\label{sec:kin}

Armed with the 1--halo components of the 3D real--space~($\xi_{1h}(r)$) and
redshift--space~($\xi^{s}_{1h}(r_{\pi}|r_p{=}0.26\,\hmpc)$; hereafter shortened as
$\xi^{s}_{1h}(r_{\pi})$) cross--correlation
functions between the LRGs and MgII absorbers, we now investigate different cloud
kinematic models by examining their predicted RSD signals and comparing to the
observations. In particular, we compute $\xi^{s}_{1h}$ by convolving $\xi_{1h}(r)$
with the LOS velocity distribution predicted by each kinematic model
\begin{eqnarray}
    \xi^{s}_{1h}(r_\pi\mid r_p)  &=& \frac{H_z}{1+z} \int_{-\infty}^{+\infty}
    \left[\xi_{1h}\big(\sqrt{r_p^2+y^2}\big)\right] \times \;\;\;\;\;\;\;\;\;\;\;\;\;\; \nonumber\\
    &&p\big(\vlos{=}\frac{H_z }{1+z} \left(r_\pi - y\right)\mid r_p, y, \boldsymbol{\theta}\big)\,\dd y,
\label{eqn:conv}
\end{eqnarray}
where $r_p{=}0.26\,\hmpc$, $p(\vlos | r_p, y, \boldsymbol{\theta})$ is the LOS
velocity distribution of cool clouds at fixed $y$, and $\boldsymbol{\theta}$ is
the parameter set of each kinematic model.

Following the discussion in~\S~\ref{sec:intro}, we consider four different kinematic
models:
\begin{itemize}
    \item Isothermal model described by a single LOS velocity dispersion $\siglos$.
    \item Satellite infall model in which the MgII absorbers reside in the CGM of
        satellite galaxies and follow the motion of satellite infall.
    \item Cloud accretion model in which clouds follow the cosmological gas accretion
        as prescribed by \citet{Afruni2019}.
    \item Tired wind model in which ancient wind bubble ``hangs'' in the outer halo,
        inspired by the semi--analytic model of \citet{Lochhaas2018}.
\end{itemize}
Among the four models, the first three are no--wind models that assume random or
inflowing motion of the clouds, while the last one assumes outflowing clouds.
Although our ultimate goal is to distinguish the four kinematic models, the current
measurement uncertainties of $\xi^{s}_{1h}$ are quite large~(${\sim}30{-}40\%$ on scales
below $3\,\hmpc$), rendering any $\chi^2$--based model comparison highly
inconclusive.  Nonetheless, we will compute the Akaike Information Criterion
(AIC) value for each best--fitting model as AIC$=2k{+}\chi^2$, where $k$ is the
number of model parameters, and compare the AIC of the four models.

More important, we will assess the capability of each model to reproduce the
sharp truncation of the observed $\xi^{s}_{1h}$ profile at
$r_{\pi}{\sim}3.4\,\hmpc$, i.e., the missing--FOG effect in the RSD
 of MgII absorbers around LRGs. Despite the large statistical uncertainties in
 individual $r_{\pi}$ bins, the
truncation feature itself is robust --- the observed $\xi^{s}_{1h}$ profile
stays roughly constant below $r_{\pi}{\sim}3\,\hmpc$, but drops rapidly from
$16.9{\pm}5.8$ at $r_{\pi}{=}3.15\,\hmpc$ to $1.2{\pm}2.6$ at
$r_{\pi}{=}3.85\,\hmpc$. We will describe each model in turn below.

\subsection{Isothermal}
\label{subsec:isothermal}

\begin{figure}
\begin{center}
    \includegraphics[width=0.48\textwidth]{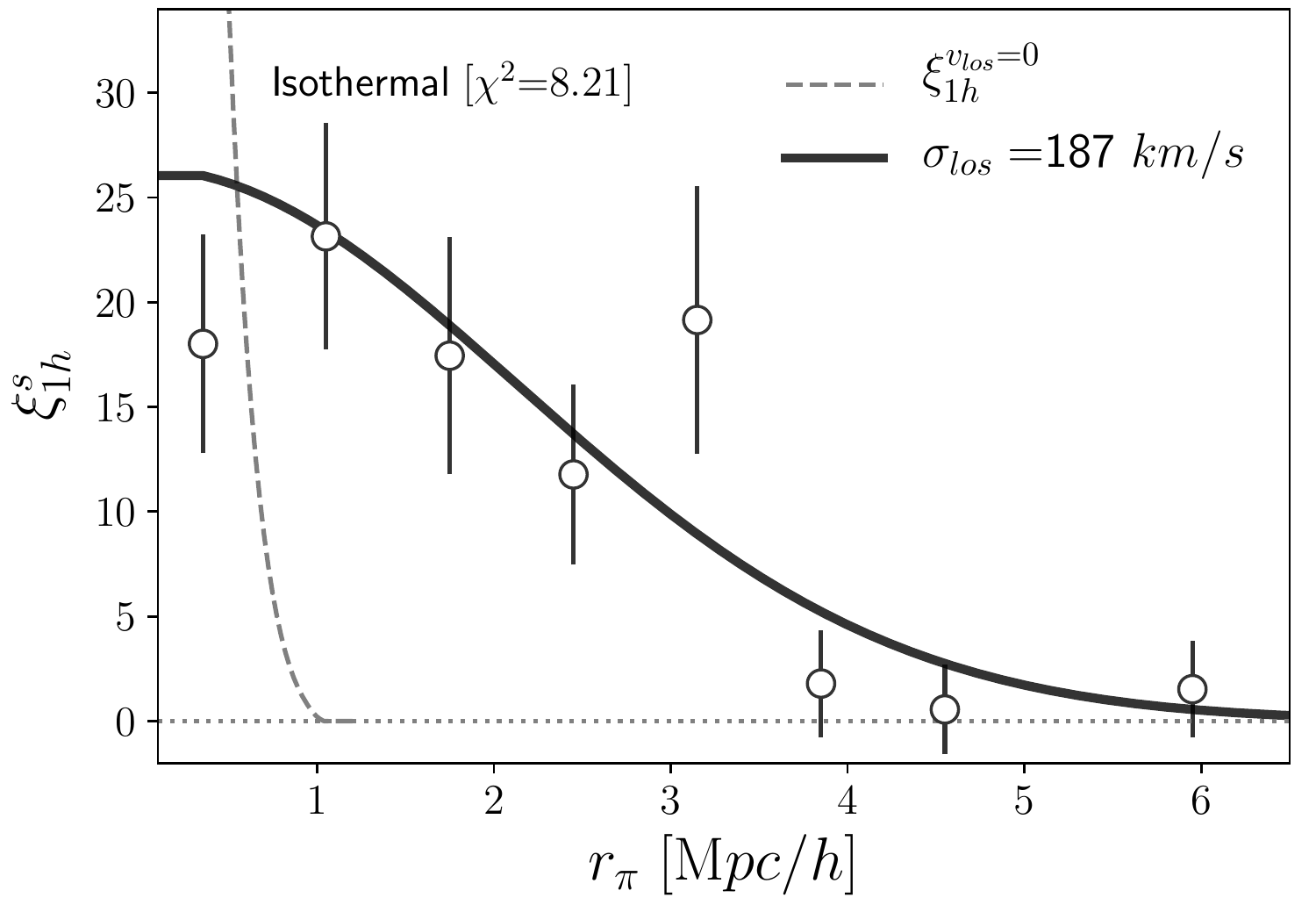}
    \caption{Comparison of the observed 1--halo components of the LRG--MgII
    absorber redshift--space cross--correlation function $\xi_{1h}^{s}$~(circles with errorbars) with the best--fitting prediction
    by the isothermal model~(solid curve). The $\chi^2$ value is indicated in the top left
    corner. Dashed line shows the predicted signal when peculiar velocity is off.}
\label{fig:isothermal}
\end{center}
\end{figure}

The isothermal model was implicitly assumed in many CGM studies, where a
velocity dispersion was usually computed from the $\Delta V$ distribution and
compared to the virial or escape velocity of the haloes. Similarly, we model
$p(\vlos)$ as a Gaussian distribution with zero mean and dispersion $\siglos$,
and predict the $\xi^{s}_{1h}$ profile by convolving $p(\vlos)$ with our
reconstructed $\xi_{1h}(r)$ profile via Equation~\ref{eqn:conv}. We derive the
best--fitting value of $\siglos$ via $\chi^2$ minimization.
Figure~\ref{fig:isothermal} compares the best--fitting prediction from the
isothermal model~(solid black curve) to the observed $\xi^{s}_{1h}$
profile~(circles with errorbars). The best--fitting value of $\siglos$ is
$187\,\kms$, lower than the virial LOS dispersion~(${\simeq}247\,\kms$ for
$\lg\,M_h{=}13.19$) and consistent with the results from previous
studies~\citep[e.g.,][]{Zhu2014, Huang2016, Anand2021, Huang2021}.  The solid
curve has a $\chi^2$ value of $8.21$, hence a statistically good fit to the
eight data points.  However, the model predicts a gradual decline of the
$\xi_{1h}^{s}$ profile, without the sharp truncation seen in the data profile.
This is unsurprising --- the isothermal model can only stretch the relative
redshift distribution of MgII absorbers around LRGs, but cannot produce a sharp
feature in the distribution.

\subsection{Satellite Infall}
\label{subsec:satellite}

\begin{figure}
\begin{center}
    \includegraphics[width=0.48\textwidth]{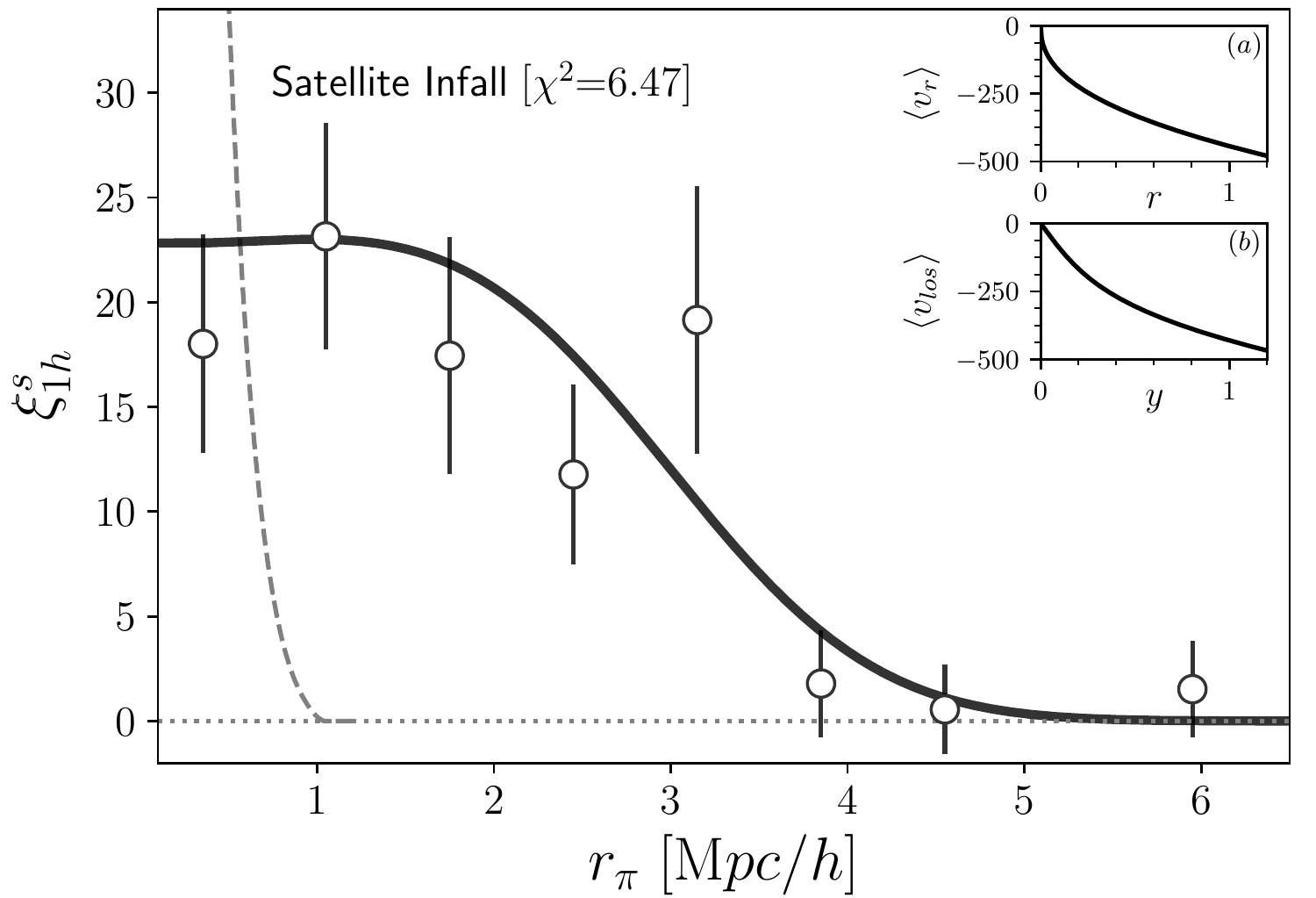}
    \caption{Similar to Figure~\ref{fig:isothermal}, but for the satellite
    infall model. Inset panels (a) and (b) indicate the average radial and LOS
    velocity profiles predicted by the best--fitting model, respectively.}
\label{fig:infall}
\end{center}
\end{figure}

Many of the LRGs are the central galaxies of group--size haloes with halo mass
above $10^{13}\,\hmsol$, which frequently accrete lower--mass systems that
likely host star-forming galaxies that could drive cool clouds into the CGM via
strong outflows. In this scenario, the observed MgII absorbers around LRGs are
actually associated with the CGM of infalling satellite galaxies. Hydrodynamic
simulations predict that it usually takes a couple Gyrs before the satellite
galaxies become quenched after being accreted onto the larger
halo~\citep{Simha2009, Wetzel2013}. Therefore, we could model the cloud
kinematics based on the velocity distribution of recently accreted satellites
inside haloes.

After examining the galaxy infall kinematics in the Millennium simulation,
\citet{Zu2013} found that the average radial velocity of the infalling galaxies
reaches maximum at some characteristic radius~(${\sim}1\,\hmpc$) and then
declines rapidly towards zero at the halo center.  Therefore, we describe the
radial velocity profile of cool clouds as
\begin{equation}
    \avg{v_r(r)} =\left( \frac{r} {1.2} \right)^\beta v_{\mathrm{max}},
\end{equation}
where $v_{\mathrm{max}}$ is the maximum radial velocity at $1.2\,\hmpc$, about
twice the $r_{200m}$ of LRG haloes.  We do not assume any rotational component
in the cloud kinematics, which is more important for disky
galaxies~\citep{Zabl2019, DeFelippis2021}.  Adopting an average projected
distance of $r_p{=}0.26\,\hmpc$, we then convert the radial velocity profile
$\avg{v_r(r)}$ to an average LOS velocity profile $\avg{\vlos(y)}$ and assume a
constant LOS velocity dispersion $\siglos$, yielding a simple satellite infall
model with three parameters $\{v_{\mathrm{max}},\, \beta,\, \siglos\}$. By
minimizing the $\chi^2$, we obtain the best--fitting parameters as
$v_{\mathrm{max}}{=}{-}479.0\,\kms$, $\beta{=}0.42$, and $\siglos{=}71.1\,\kms$,
respectively.

Figure~\ref{fig:infall} shows the best--fitting prediction from our satellite
infall model~(solid black curve), with the two inset panels indicating the
best--fitting average radial~(top) and LOS~(bottom) velocity profiles,
respectively.  Compared to the isothermal model~($k{=}1$), the minimum $\chi^2$
decreases from $8.21$ to $6.47$ with two more parameters~($k{=}3$). On the one
hand, based on the classical Akaike information criterion~(AIC$=2k+\chi^2$), the
satellite infall model has a slightly higher AIC value~($12.47$) than the
isothermal model~($10.21$), hence a slightly less preferred model; But on the
other hand, the satellite infall model produces a more flattened profile on
small scales and a steeper decline, hence a better match to the shape of the
$\xi^{s}_{1h}$ profile than the isothermal model. However, our satellite infall
model still cannot reproduce the observed sharp truncation at
$r_{\pi}{\sim}3.4\,\hmpc$.

\subsection{Cloud Accretion}
\label{subsec:accretion}

\begin{figure}
\begin{center}
    \includegraphics[width=0.48\textwidth]{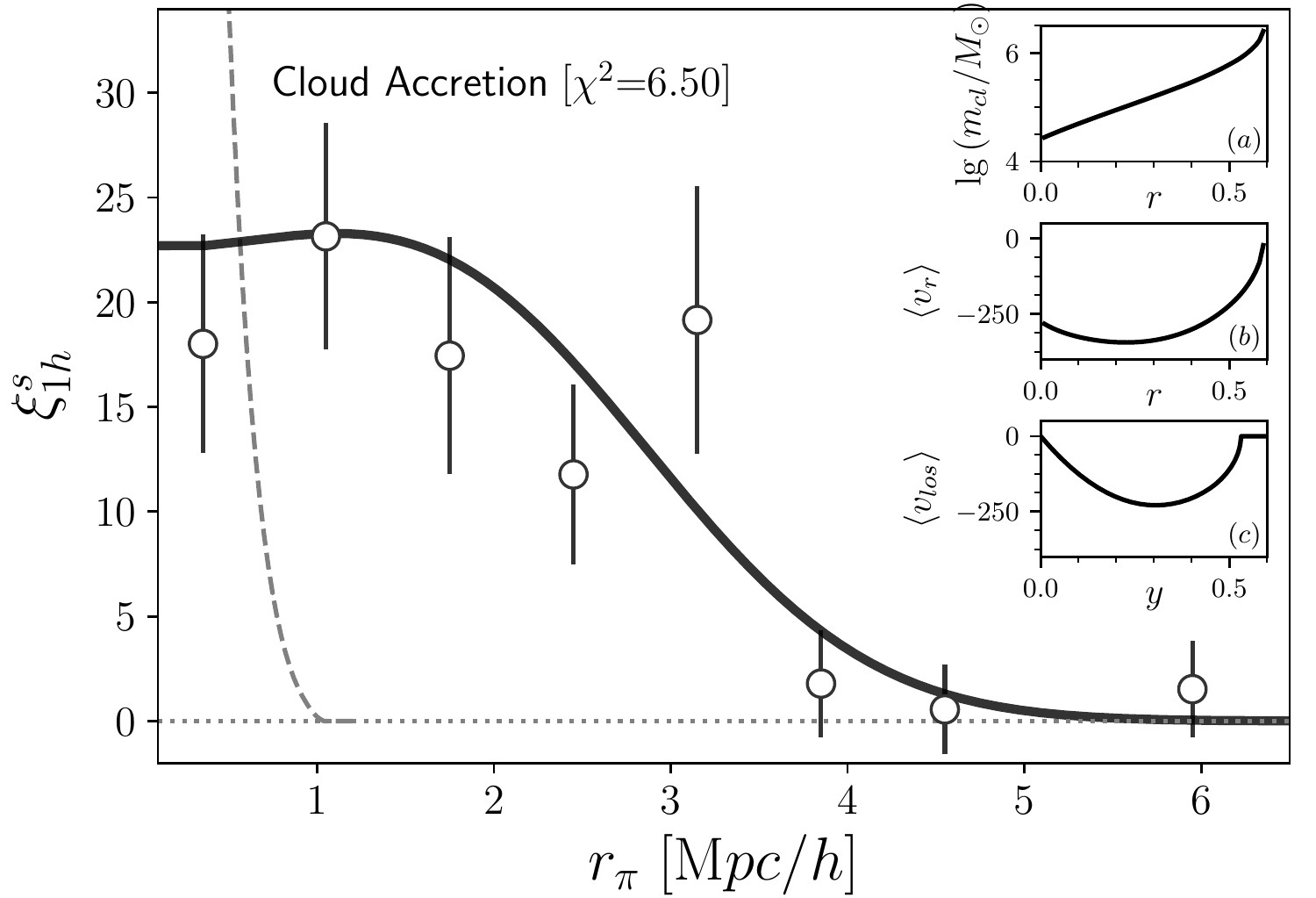}
    \caption{Similar to Figure~\ref{fig:isothermal}, but for the cloud accretion model.
    Inset panels (a), (b), and (c) show the cloud mass evaporation, average radial velocity
    profile, and average LOS velocity profile, respectively.}
\label{fig:afruni}
\end{center}
\end{figure}

Apart from infalling with the satellite galaxies, the observed cool clouds could
also be part of the cosmological inflow that sustains the growth of LRG haloes.
We adopt the same semi--analytic model of cloud accretion developed
\citet{Afruni2019} to predict the radial and LOS velocity profiles of gas
clouds; But instead of assuming a
theoretical cool gas accretion rate as in \citet{Afruni2019}, we convolve the
LOS velocity distribution with our reconstructed $\xi_{1h}(r)$ profile to
predict the $\xi^{s}_{1h}$ profile via Equation~\ref{eqn:conv}. In this way,
our comparison with the observation is self--consistent, because both
$\xi_{1h}(r)$ and $\xi^{s}_{1h}$ are derived from observations and the cloud
kinematics is the only unknown in the problem. We will briefly describe the
relevant components of the \citet{Afruni2019} model below and refer the reader
to \citet{Afruni2019} for more technical details.

The \citet{Afruni2019} model describes the radial
equation of motion of the cool clouds as
\begin{equation}
    \frac{\dd v_r}{\dd r} = \frac{1}{v_r(r)} \frac{GM(<r)}{r^2} - \frac{\pi\,r^2_{cl}\,
    \rho_{\mathrm{cor}}(r)\, v_r(r)}{m_{cl}(r)},
    \label{eqn:eom}
\end{equation}
where $M(<r)$ is the halo mass enclose within $r$, $\rho_{\mathrm{cor}}$ is the
density of the hot corona gas, and $r_{cl}$~($m_{cl}$) is the radius~(mass) of a
typical gas cloud. The second term on the r.h.s. of Equation~\ref{eqn:eom} describes the
deceleration of the cool clouds due to ram pressure drag, and $\rho_{\mathrm{cor}}$ is
derived from the NFW potential by assuming hydrostatic equilibrium of the hot gas
with the dark matter halo. The mass and radius of the gas cloud are related via
\begin{equation}
    r_{cl} = \left(\frac{3 m_{cl}}{4 \pi \rho_{cl}}\right)^{1/3},
    \label{eqn:rcl}
\end{equation}
where $\rho_{cl}$ is the density of the cool clouds, calculated from assuming
pressure equilibrium between the cool and hot phases of the CGM. Additionally, the mass
loss due to hydrostatic instabilities~\citep{Armillotta2017} is modelled as
\begin{equation}
    \frac{\dd\,m_{cl}}{\dd\,r} = - \alpha \frac{m_{cl}(r)}{v_r(r)},
    \label{eqn:evap}
\end{equation}
where $\alpha$ is the cloud evaporation rate. Following \citet{Afruni2019}, we
assume all the clouds start with zero velocity at $r_{200m}{=}0.6\,\hmpc$ and
initial cloud mass $m_{0,cl}$, and then glide through an LRG halo with
$\lg\,M_h{=}13.19$. By combining Equations~\ref{eqn:eom}, \ref{eqn:rcl}, and
\ref{eqn:evap} together, we can solve for the radial velocity profile
$\avg{v_r(r)}$ of cool clouds for any given values of $\alpha$ and $m_{0,cl}$,
hence the LOS velocity profile $\avg{\vlos(y)}$ at $r_p{=}0.26\,\hmpc$.
Finally, assuming a constant LOS velocity dispersion $\siglos$, we can predict
the $\xi^{s}_{1h}$ profile from $\xi_{1h}(r)$ using Equation~\ref{eqn:conv}.  By
minimizing the $\chi^2$, we find the best--fitting parameters of the cloud
accretion model to be $\alpha{=}2.10\,\mathrm{G}yr^{-1}$, $\lg\,m_{0,
cl}{=}6.47$, and $\siglos{=}98.5\,\kms$, respectively. Compared to the
constraints from \citet{Afruni2019}, our mass loss rate is similar to
theirs~($\alpha{=}1.8\,\mathrm{G}yr^{-1}$), but our cloud mass is larger by a
factor of 30~($\lg\,m_{0, cl}{=}4.85$). The cloud mass discrepancy is
unsurprising because we rely only on the kinematics to infer $m_{cl}$ while
\citet{Afruni2019} used both the velocity distribution and the hydrogen column
densities.


Figure~\ref{fig:afruni} shows the best--fitting prediction from the cloud
accretion model~(solid black curve), with the three inset panels indicating the
best--fitting average mass evaporation~(top), radial velocity~(middle), LOS
velocity~(bottom) profiles, respectively. Despite significant differences in the
predicted radial velocity profiles, the cloud accretion model is almost
indistinguishable from the satellite infall model, with nearly the same values
of minimum $\chi^2$~($6.47$ vs. $6.50$) and the same number of parameters~(hence
similar AIC values $12.47$ vs. $12.50$). Meanwhile, similar to the satellite
infall model, the cloud accretion model produces a flattened small--scale profile followed
by a decline that is less sharp than seen in the data profile.

\subsection{Tired Wind}
\label{subsec:wind}

\begin{figure}
\begin{center}
    \includegraphics[width=0.48\textwidth]{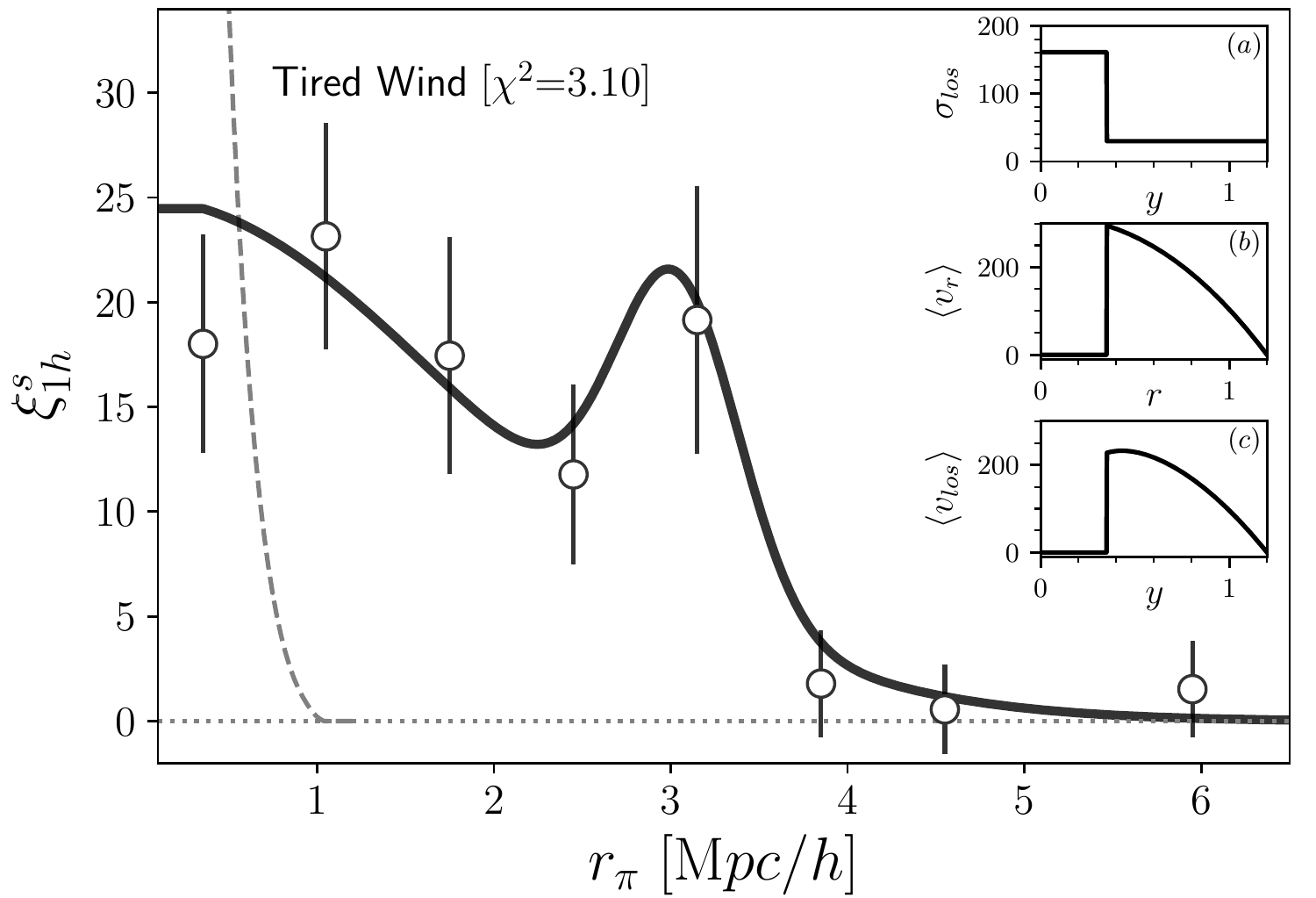} \caption{Similar to
    Figure~\ref{fig:isothermal}, but for the tired wind model. Inset panels (a), (b), and
    (c) indicate the LOS velocity dispersion, average radial velocity, and average LOS
    velocity profiles, respectively.}
\label{fig:outflow}
\end{center}
\end{figure}

The three no--wind models we have explored so far could all provide
statistically good fits to the RSD data, but are unable to reproduce the sharp
truncation in the observed $\xi^{s}_{1h}$ profile. In particular, the two
inflow--based models could produce a similarly flattened profile on small
scales, but the decline at $r_{\pi}{\sim}3{-}4\,\hmpc$ is not as steep as in the
data profile. An outflow--based model may be able to obtain a better match by
propagating a wind bubble to the outer CGM, as predicted by the model of
~\citet{Lochhaas2018}.

To mimic the kinematic features of ancient wind bubbles driven by the early star
formation of LRGs, we adopt a simple ``tired wind'' prescription consisting of a
dispersion--dominated component at small radii and a radially outflowing
component at large radii. In particular, the radial velocity profile is
\begin{eqnarray}
    \avg{v_r(r)} &=&   \left\{
\begin{array}{ll}
 0 & \quad\mbox{if $r < r_{\mathrm{min}} $ },\nonumber\\
    \left(1 -\left(r/1.2\right)^\gamma\right) v_0 & \quad\mbox{if $r \geqslant
    r_{\mathrm{min}} $ },\nonumber
\end{array}
\right.\\
\label{eqn:outflow}
\end{eqnarray}
where $v_0$ is the launch velocity of the wind bubbles from the halo center and
$r_{\mathrm{min}}$ is the minimum radius of the radiatively cooled wind bubble,
driven by the last star formation episode of the LRGs before quenching. On
scales below $r_{\mathrm{min}}$, we assume random motion with LOS velocity
dispersion ${\sigma}_{\mathrm{los}}$ for the cool clouds, which could be formed
via shell fragmentation due to various instabilities; On scales above
$r_{\mathrm{min}}$, we assume a constant LOS velocity dispersion of $30\,\kms$
for the outflow component. This choice of dispersion is small enough so that
each wind bubble remains a coherent thin--shell without fragmentation, and large
enough to avoid overfitting the data. Therefore, we have in total four parameters:
$r_{\mathrm{min}}$, $\siglos$, $v_0$, and $\gamma$, and their best--fitting values from
$\chi^2$ minimization are $0.35\,\hmpc$, $161.0\,\kms$, $317.0\,\kms$, and $2.18$,
respectively.

Figure~\ref{fig:outflow} shows the best--fitting prediction from the tired wind
model~(solid black curve), with the three inset panels indicating the
best--fitting LOS velocity dispersion~(top), radial velocity~(middle), and LOS
velocity~(bottom) profiles, respectively.  The tired wind model provides an
excellent description to the data with $\chi^2{=}3.1$ and a reasonable AIC
value of $11.10$. Clearly, the significant improvement in $\chi^2$ from the
other three models is mainly driven by the better match to the observed
truncation of $\xi^{s}_{1h}$, thanks to the survived wind bubble moving at
$250\,\kms$ at $r{\sim}400\,\hkpc$. This tired wind scenario is also roughly
consistent with the theoretical expectation from the \citet{Lochhaas2018} model.
While the \citet{Lochhaas2018} model tracks the radial motion of a single
wind--bubble as a function of time for an individual galaxy, our radial velocity
profile effectively describes the synthetic snapshot of a continuous series of
wind--bubbles launched at different epochs by {\it different} galaxies.

Thus far, all the four best--fitting models we explored provide statistically
good fits to the RSD of MgII absorbers around LRGs, yielding similar AIC values
that render no particular model more preferable than the other three. However,
if we focus on reproducing the truncation feature in the observed
$\xi^{s}_{1h}$, the tired wind model clearly stands out, by interpreting the
truncation as the kinematic signature of ancient wind bubbles.

\section{Summary and Look to the Future}
\label{sec:conc}

In this paper, we have measured the LRG--MgII absorber cross--correlation
function in the redshift space and projected along the LOS, $\xirs(r_p,
r_{\pi})$ and $w_p(r_p)$, respectively.  Compared to the LRG auto--correlation,
$\xirs(r_p, r_{\pi})$ exhibits a similar Kaiser effect on large scales, but a
much weaker FOG effect on small scales. In particular, the $\xi^{s}(r_{\pi})$
profile measured at $r_p{=}0.26\,\hmpc$ is heavily truncated at
$r_{\pi}{\sim}3.4\,\hmpc$, likely due to a nontrivial feature in the motion of
cool clouds around the LRGs.

Combining $w_p(r_p)$ and the stacked MgII absorption profile measured
by~\citet{Zhu2014}, we successfully reconstructed the 3D isotropic
cross--correlation function $\xi(r)$ between the MgII absorbers and LRGs. We
found that the detection completeness of the resolved MgII absorbers can be
described by a simple power--law function of distance away from the
LRGs~(${\sim}r^{1.2}$), yielding a flat covering fraction profile despite the
stacked MgII absorption follows a projected NFW profile.  Furthermore, we
extracted the 1--halo component of the 3D correlation function $\xi_{1h}(r)$,
which provides us the real--space LOS distribution of cool clouds within the LRG
haloes, i.e., $\xi_{1h}(y{=}(r^2-r_{p}^2)^{1/2})$ at $r_p{=}0.26\,\hmpc$.

To obtain the redshift--space LOS distribution of cool clouds within the LRG
haloes, we have estimated the 1--halo component of the $\xi^{s}(r_{\pi})$
profile of MgII absorbers at $r_p{=}0.26\,\hmpc$. In particular, we subtracted
the 2--halo component empirically derived from the $\xi^{s}(r_{\pi})$ profile
at $r_p{=}0.75\,\hmpc$, i.e., externally tangent to the LRG haloes. The sharp
truncation feature seen in $\xirs$ persists in the estimated 1--halo term
$\xi^{s}_{1h}(r_{\pi})$ at $r_p{=}0.26\,\hmpc$, which can be compared with
predictions from convolving different kinematic models of the cool clouds with
$\xi_{1h}(y)$.

We consider four kinematic models~(three without wind and one with wind) of the
MgII absorbers, including an isothermal model in which cool clouds move randomly
with a single LOS velocity dispersion, a satellite infall model in which cool
clouds are associated with the infalling satellites, a cloud accretion model in
which cool clouds join the cosmological inflow of dark
matter~\citep{Afruni2019}, and a tired wind model in which cool clouds originate
from the fragmentation and propagation of ancient wind
bubbles~\citep{Lochhaas2018}. For each model, we derived the best--fitting
parameters by minimizing the $\chi^2$ between the observed and predicted
$\xi^{s}_{1h}(r_{\pi})$ at $r_p{=}0.26\,\hmpc$. Based on the $\chi^2$ values,
all the four models provide statistically good fits to the data. They also yield
similar values of AIC, suggesting that the current measurement uncertainties are
too large to statistically distinguish between the no--wind vs. wind models.

Interestingly, we found that only the tired wind model is capable of reproducing
the observed sharp truncation in the redshift--space LOS distribution of MgII
absorbers within the LRG haloes, while the other three fail to do so. The tired
wind model explains the truncation with an ancient wind bubble expanding at
${\sim}250\,\kms$ on scales around $400\,\hkpc$. This physical picture
is roughly consistent with the theoretical expectation from the
\citet{Lochhaas2018} model, in which they discovered that galactic wind--driven
bubble could cool radiatively and then ``hang'' in the outer halo for several
Gyrs.

Although our finding supports the tired wind scenario, it is far from being
conclusive. Due to the large uncertainties in the measurement of $\xirs$, our
results are also consistent with both the dispersion and inflow--dominated cloud
kinematics. However, we have convincingly demonstrated the efficacy of our
method for inferring cool cloud kinematics from the RSD of MgII absorbers. In
particular, the cloud kinematics is the missing link between the projected and
redshift--space cross--correlation functions between the MgII absorbers and
galaxies, and our method is able to exploit the spherical symmetry in
cross--correlation analysis to infer such missing link in a self--consistent
manner. With ongoing and future spectroscopic surveys like the DESI and PFS, we
will obtain a greater number of quasar sightlines, hence greater number of MgII
absorbers~\citep{Zhao2019, Anand2021}, and apply the method to massive quiescent
galaxies as well as a large number of emission line galaxies.

\section*{Data availability}

The data underlying this article will be shared on reasonable request to the corresponding author.

\section*{Acknowledgements}

We thank Cassi Lochhaas, Ting-Wen Lan, and David Weinberg for helpful
discussions, and the anonymous referee for comments that have greatly improved
the manuscript. YZ acknowledges the support by the National Key Basic Research
and Development Program of China (No.  2018YFA0404504), National Science
Foundation of China (11873038, 11621303, 11890692), the science research grants from the
China Manned Space Project (No. CMS-CSST-2021-A01, CMS-CSST-2021-B01), the National One-Thousand
Youth Talent Program of China, and the SJTU start-up fund (No. WF220407220). YZ
thanks the stimulating discussions with Cathy Huang during his quarantine at the
Zhangjiang Hi-Tech Park during the pandemic.






\bsp	
\label{lastpage}

\end{document}